\documentclass[a4paper,11pt]{article}
\pdfoutput=1 

\usepackage{jheppub} 

\usepackage[T1]{fontenc} 

\usepackage[utf8]{inputenc}
\usepackage{mathtools,amsmath,amsfonts,amsthm,amssymb}
\usepackage{tikz-cd}
\usepackage{adjustbox}
\usepackage{stackrel}
\usepackage{graphicx}
\usepackage{float}
\usepackage{pst-node}
\usepackage{etoolbox}
\usetikzlibrary{matrix,arrows}
\definecolor{pur}{RGB}{186,146,162}
\usepackage{pdfpages}

\usepackage{currvita}

\usepackage[framemethod=tikz]{mdframed}

\title{\boldmath Bosonization based on Clifford algebras\\and its gauge theoretic interpretation}

\author{A. Bochniak and B. Ruba}

\affiliation{Institute of Theoretical Physics, Jagiellonian University in Krak\'{o}w,\\ prof. {\L}ojasiewicza 11, 30-348 Kraków, Poland}

\emailAdd{arkadiusz.bochniak@doctoral.uj.edu.pl}
\emailAdd{blazej.ruba@doctoral.uj.edu.pl}

\abstract{We study the properties of a bosonization procedure based on Clifford algebra valued degrees of freedom, valid for spaces of any dimension. We present its interpretation in terms of fermions in presence of $\mathbb Z_2$ gauge fields satisfying a modified Gauss' law, resembling Chern-Simons-like theories. Our bosonization prescription involves constraints, which are interpreted as a flatness condition for the gauge field. Solution of the constraints is presented for toroidal geometries of dimension two. Duality between our model and $(d-1)$-form $\mathbb Z_2$ gauge theory is derived, which elucidates the relation between the approach taken here with another bosonization map proposed recently.}

\begin{document} 
\maketitle
\flushbottom

\section{Introduction}

Many fermionic systems admit bosonizations, i.e.\ alternative descriptions formulated using bosonic operators. Such correspondences are especially abundant for theories formulated in spacetime dimension two \cite{jordan,witten,senechal}. Their importance stems from the fact that they allow to construct analytic solutions of certain models \cite{lieb, mandal}, to gain nonperturbative insights into dynamics of strongly coupled systems \cite{strongly} and, more recently, to understand certain phases of topologically nontrivial fermionic matter \cite{kt}. Furthermore, there exist systems for which dualities help to overcome problems in numerical studies, such as the sign problem in Monte Carlo simulations \cite{detar, dgs2013, gks2015} or difficulties in implementation of operators acting on Hilbert spaces which do not factorize into tensor products of on-site Hilbert spaces. This last problem may also have some significance for the field of quantum information \cite{kitaev,kitaev1}.

The most well-known bosonization methods apply only to $1+1$-dimensional systems. Some proposals valid in higher dimensions have been put forward \cite{ckr, ck, W, kt,burgess,kopietz, bravyi, ball, verstraete,fradkin, ktong, zohcir}. See also reviews in \cite{tong, Th19, Son18, web}. Each of these constructions involves some difficulties not present for two-dimensional systems, such as non-locality or presence of complicated constraints, often interpreted as the Gauss' law of some gauge theory. One~could argue that this is an~inherent feature of models involving fermionic degrees of freedom. Further study of these phenomena might help to eventually construct bosonization maps more suitable for practical calculations, which is the main motivation of this work.

The main part of this paper is concerned with the study of the bosonization method proposed in \cite{W}. In~this approach fermion fields are replaced by on-site Euclidean $\Gamma$ matrices. For~this reason we call it the $\Gamma$ model. This model is bosonic in the sense that the $\Gamma$ matrices, which serve as its elementary fields, commute when placed on distinct lattice sites. Moreover its~Hilbert space is the tensor product of Hilbert spaces associated to individual lattice sites. The~price to pay for this convenience is the necessity to introduce certain constraints on physical states. Correspondence between the $\Gamma$ model with constraints and fermions, at least for the free fermion hamiltonian, has been conjectured based on a~comparison between relations satisfied by operators present in hamiltonians of these two models. Precise statement of this correspondence has been formulated and proven for the first time in \cite{szczerba}. It~turned out that the proposed bosonization map is valid for any hamiltonian, hence purely kinematical. Here we extend it by considering more general geometries. We~provide a new proof of validity of this construction, inspired by techniques from \cite{ckr}. Furthermore, we provide a new interpretation of constraints present in the $\Gamma$ model as the pure gauge condition for a certain $\mathbb Z_2$ gauge field. We show that fermions coupled to general $\mathbb Z_2$ gauge fields can be modeled by modifying the form of constraints, without altering the form of the bosonized hamiltonian. The full Hilbert space of the $\Gamma$ model decomposes into a direct sum of subspaces corresponding to all possible gauge fields. This decomposition has the interesting property that only states with specific fermionic parity, depending on the gauge field, are present. We illustrate the main features of our model by presenting examples in the cases of a specific geometry (two-dimensional tori, for which we also solve the constraints) and for a simple class of solvable fermionic hamiltonians. This work parallels \cite{brww}, which motivated our studies, allowed to formulate initial hypotheses and test them using symbolic algebra software.

It is natural to ask whether it is possible to make the gauge field present in the $\Gamma$ model dynamical. In other words, does the $\Gamma$ model with no constraints imposed provide a bosonization of a some theory of fermions coupled to a~$\mathbb Z_2$ gauge field? We show that such mapping does indeed exist. It is local for even fermionic operators and for gauge field operators of magnetic type\footnote{We call an operator magnetic if it is a function of the gauge field on a~single time slice and electric if it acts by flipping the gauge field. General observables in gauge theory involve operators of both types.}, but operators involving the electric field are represented in a~complicated way, which depends on a~choice of a loop wrapping around the whole lattice. Similarly, the elementary field of the $\Gamma$ model is non-local on the gauge theory side.

Gauge theory corresponding to the unconstrained $\Gamma$ model involves a mechanism present in the Dijkgraaf-Witten theory \cite{dwitten,freed,wan_wang_he} and more general gauge theories with Chern-Simons-like topological terms: Hilbert space representation of time-independent gauge transformations, here written for simplicity in the $\mathrm{U}(1)$ continuum theory language\footnote{Precise formulation suitable for our lattice models is given in the main text.},
\begin{equation}
| A_{i} \rangle \mapsto | A_{i} + \partial_{i} \theta \rangle
\end{equation}
are modified by introducing gauge field dependent phase factors:
\begin{equation}
| A_{i} \rangle \mapsto e^{i I(\theta, A)} | A_{i} + \partial_{i} \theta \rangle.
\label{eq:I_functional}
\end{equation}
This has the consequence that the Gauss' law is altered, which leads to a deformation of the algebra of gauge-invariant operators. In particular, the constraint on the total charge, obtained by integrating the Gauss' law over the whole space, is modified. This is the celebrated flux attachment mechanism \cite{wilczek}: electric excitations in models of this type are decorated by magnetic fields. Braiding of two such excitations involves Aharonov-Bohm phases, leading to a transmutation of statistics. In our case, the total number of fermions modulo two becomes related to the value of a~certain magnetic observable. An unpleasant feature of the gauge theory corresponding to the $\Gamma$ model is that the functional $I(\theta,A)$ in \eqref{eq:I_functional} depends non-locally on the gauge field $A$. We demonstrate that under certain assumptions about the lattice this non-locality may be removed by a~canonical transformation which preserves the form of all fermionic and magnetic observables (so~bosonization is still local for those operators for which it initially was).

There exists a duality mapping which relates the $\Gamma$ model to higher gauge theories proposed in the context of bosonization in \cite{ckr, ck, chen}. In some aspects it resembles the classical Kramers-Wannier duality \cite{kramers}. It is clear that this correspondence has to involve a transition to the dual spatial lattice. Indeed, in our model local degrees of freedom act on Hilbert spaces associated to lattice sites, just as in the initial fermionic theory, while constraint operators are located on plaquettes. In the latter case, for spacetimes of dimension $d+1$, degrees of freedom associated to $(d-1)$-cells have been proposed, with fermionic operators placed on $d$-simplices and constraints on $(d-2)$-simplices. This setup has the advantage that it is naturally interpreted in terms of $(d-1)$-form gauge theory (involving the flux attachment mechanism). On the other hand, our formulation is more uniform, in the sense that it applies in unchanged form in any dimension. The amount of redundancy in the two approaches (defined as the ratio of the dimension of the full Hilbert space and the subspace defined by constraints) is the same order (and rather large) in both cases. Secondly, in our construction it is crucial that each lattice vertex is incident to an even number of edges. We~remark here that it is possible to define the $\Gamma$ model even if this condition is not satisfied, but in this case it is found to contain additional degrees of freedom, resembling Majorana fermions. This feature is discussed in the appendix \ref{sec:odd_degree}. 

The organization of this paper is as follows. In section \ref{sec:geometric_seput} we recall basic geometric concepts used in the main text. Reader not at all familiar with this language may want to consult introductory books in algebraic topology (see e.g.~\cite{hatcher}) first. Section \ref{sec:fermions} is concerned mainly with the review of a known description of the algebra of even fermionic operators in terms of a~convenient set of generators and relations. The main part of the text starts in section \ref{sec:gamma_model}. In subsections \ref{sec:gamma_def}, \ref{sec:rep_ambig} we define the $\Gamma$ model and establish its correspondence with fermions. Then we derive the gauge-theoretic interpretation of this model in subsection~\ref{sec:modified_constraints}. Presented constructions are illustrated by the example of toroidal geometry, discussed in the subsection \ref{sec:toroidal} and the discussion of quadratic hamiltonians in \ref{sec:quadratic}. In~the special case of dimension $2+1$ we solve the constraints relevant for our bosonization procedure and relate them to ground states of the Kitaev's toric code~\cite{kitaev}. Section \ref{sec:deformed_gauss} is devoted to the study of modified gauge theories. Proof of the equivalence between the gauge theory proposed in the subsection \ref{sec:modified_constraints} and the $\Gamma$ model is presented in the subsection \ref{sec:gauge_invariant_ops}. Afterwards a~generalization of this gauge model, involving modified Gauss' operators, is~introduced in the subsection \ref{sec:classifaction}. We~classify these theories up to equivalence given by (in~general non-local) canonical transformations. This allows to find a local formulation of the gauge theory corresponding to the $\Gamma$ model in the subsection \ref{sec:local}. Afterwards, in section \ref{sec:higher_gauge}, we present the duality between the $\Gamma$ model and higher gauge theory. This includes a brief discussion of the role of spin structures. We summarize in section~\ref{sec:summary}. The~paper is closed with two appendices. Appendix \ref{canonical_heisenberg} is concerned with Heisenberg groups and their automorphisms for $\mathbb Z_2$-valued degrees of freedom, while appendix \ref{sec:odd_degree} discusses the extension of the $\Gamma$ model to the case in which some vertices are incident to an~odd number of edges.

\section{Geometric setup\label{sec:geometric_seput}}
For any finite set $S$ we let $|S|$ be the number of elements of $S$.

All physical systems will be considered on a connected graph $\mathfrak G=(V,E)$, which may (but~does not have to) be the set of vertices and edges of a triangulation or more general cell decomposition of some manifold. We will assume that the graph $\mathfrak G$ is such that every edge connects two distinct vertices. Multiple edges which connect the same vertices are allowed. We let $E_{\mathrm{or}}$ be the set of oriented edges. Thus every edge $ e \in E$ corresponds to two distinct elements of $E_{\mathrm{or}}$. We~have functions $s,t : E_{\mathrm{or}} \to V$, called source and target maps, which assign to $e \in E_{\mathrm{or}}$ its initial and final vertex, respectively. Furthermore, for every $e \in E_{\mathrm{or}}$ we let $\overline e$ be the same edge with its orientation reversed, so that $s(\overline e) = t(e)$ and $t(\overline e)=s(e)$. If $v=s(e)$ or $v=t(e)$, we say that $e$ contains $v$ and write $v\in e$. The star $\mathrm{St}(v)$ of a vertex $v \in V$ is defined as the set of all $e \in E$ which contain $v$. Number $\mathrm{deg}(v):= |\mathrm{St}(v)|$ is called the degree of $v$.

In order to keep track of various signs we shall use the language of chains, which are formal sums of geometric objects with coefficients in the field $\mathbb Z_2$ (integers modulo~$2$). More precisely, $C_0$ and $C_1$ are defined as the $\mathbb Z_2$-vector spaces with bases $V$ and~$E$, respectively. Linear map $\partial : C_1 \to C_0$, called the boundary operator, is defined first on basis elements by $\partial e = \sum\limits_{v \in e} v$. Its kernel (called the set of cycles) and image (called the set of boundaries) are denoted by $Z_1$ and $B_0$, respectively. There are perfect bilinear pairings $C_p \times C_p \to \mathbb Z_2$, given by $(v,v') = \delta_{v,v'}$ and $(e,e')=\delta_{e,e'}$. This allows to identify chain groups $C_p$ with cochain groups $C^p := \mathrm{Hom}(C_p,\mathbb Z_2)$. Coboundary operator $C^0 \to C^1$ is defined as the adjoint of $\partial$, i.e.~by $(\delta \epsilon, \tau) = (\epsilon, \partial \tau)$ for $\epsilon \in C^0$ and $\tau \in C_1$. Equivalently, $\delta v = \sum\limits_{v \in e} e$. Kernel and image of $\delta$ are denoted by $Z^0$ and $B^1$ and called the set of cocycles and the set of coboundaries, respectively. By~construction, cocycles are orthogonal to boundaries, while coboundaries are orthogonal to cycles. In particular, there is an induced non-degenerate pairing $Z_1^\ast \times Z_1 \to \mathbb Z_2$, where $Z_1^\ast := C^1/B^1$. Thus $Z_1^\ast$ may be identified with the dual space of $Z_1$. The image in $Z_1^\ast$ of an element of $A \in C^1$ will be denoted by $[A]$. 

For future reference we calculate the dimension of $Z_1$ (and hence also of $Z_1^\ast$) over $\mathbb Z_2$. In~general the dimension of the domain of a linear operator is the sum of dimensions of the kernel and the range. Applying this to $\partial$ we obtain $\dim (Z_1) = \dim(C_1) - \dim(B_0)$. Connectedness of $\mathfrak G$ means that $\dim(B_0) = \dim(C_0) -1$. Therefore
\begin{equation}
\dim(Z_1) = \dim(C_1) - \dim(C_0) + 1 = |E| - |V| +1.
\label{eq:Z1_dim}
\end{equation}
This means that each of sets $Z_1$ and $Z_1^*$ has $2^{|E|-|V|+1}$ elements.

Tuple of oriented edges $\ell = (e_1,...,e_n)$ will be called a path if $t(e_i)=s(e_{i+1})$ for $i<n$. We will say that $\ell$ is a circuit if $t(e_n) = s(e_1)$. For every path $\ell$ we let $[ \ell ] = \sum\limits_{i=1}^n e_i \in C_1$, where we forget the orientations of $e_i$. Chain $[ \ell ]$ is a cycle if and only if $\ell$ is a circuit. Circuit $\ell$ is said to be Eulerian if every edge $e \in E$ occurs exactly once among $e_1,...,e_n$. For every such circuit we have $[ \ell ] = \sum \limits_{e \in E} e $. It is a classical result \cite[section 4.2.1]{graph} in graph theory that Eulerian circuit exists if and only if every vertex has even degree. Clearly, the latter condition is equivalent to closedness of the chain $\zeta := \sum \limits_{e \in E} e \in C_1$, i.e.\ to $\partial \zeta =0$. 

In some parts of this work (not essential for the main construction) we will have to assume that besides vertices and edges, the considered lattice is also equipped with a set of faces $F$, which are polygons whose sides are identified with edges. This allows to define the space of $2$-chains $C_2$ with an obvious boundary map $\partial : C_2 \to C_1$. Its kernel and image are denoted by $Z_2$ and $B_2$, respectively. Homology group $H_1$ is defined as the quotient $Z_1/ B_1$. There is also a scalar product $C_2 \times C_2 \to \mathbb Z_2$ given by $(f,f') = \delta_{f,f'}$ for $f,f' \in F$. Dualizing, there is also a coboundary map $\delta : C^1 \to C^2$ with kernel and image $Z^1, B^2$. Cohomology group $H^1=Z^1/B^1$ is the dual space of $H_1$.

\section{Fermions - generators and relations} \label{sec:fermions}

Here we consider a specific class of fermionic models, defined below. We~emphasize those properties that are used to prove validity of our bosonization prescription. In particular, we describe the algebra of~even fermionic operators in terms of generators and relations. This result is similar to one in \cite{ckr}, with the statement and the proof adjusted to the fact that we work with finite, not~necessarily simply-connected lattices. Our considerations are independent of dynamics, so we do not focus on any particular hamiltonian. In most of this section we repeat well-known facts, to some extent to fix notation.

First, let us denote by $\mathcal{A}$ the complex $\ast$-algebra generated by elements $\phi^\ast(v)$ and $\phi(v)$ (called creation and annihilation operators located at the vertex $v$) with $v\in V$, subject to the canonical anticommutation relations
\begin{equation}
\{ \phi(v), \phi(v') \} = \{ \phi^{\ast}(v), \phi^\ast(v') \} = 0, \qquad \{ \phi(v) , \phi^\ast(v') \} = \delta_{v,v'}.
\label{eq:Fermi_algebra_relations}
\end{equation}
By construction, every element of $\mathcal A$ may be written down as a linear combination of products of creation and annihilation operators. It is often useful to use a different set of generators of $\mathcal A$, e.g.\ the so-called Majorana operators:
\begin{equation}
    X(v)=\phi(v)+\phi^\ast(v),\qquad Y(v)=i(\phi(v)-\phi^\ast(v)).
\end{equation}
Defining relations \eqref{eq:Fermi_algebra_relations} are equivalent to 
\begin{equation}
\{ X(v), Y(v') \}=0, \qquad \{ X(v), X(v') \} = \{ Y(v), Y(v') \} = 2 \delta_{v,v'}.
\end{equation}
This shows that $\mathcal{A}$ is a Clifford algebra on $2|V|$ generators, and hence it is isomorphic to $\mathrm{End}(\mathcal F)$, the algebra of linear operators on the unique (up to isomorphism) irreducible representation $\mathcal F$ of $\mathcal A$. Dimension of $\mathcal F$ is equal to $2^{|V|}$. Every finite-dimensional representation of $\mathcal A$ is a~direct sum of finitely many copies of the irreducible representation.

Representation $\mathcal F$ is, of course, the Fock space. It is a Hilbert space with a distinguished element $| 0 \rangle$ (called the vacuum state), determined uniquely up to phase by the conditions $\phi(v) | 0 \rangle =0$ and $\langle 0 | 0 \rangle =1$. Other states, labeled by $\mathbb Z_2$-valued $0$-chains $\epsilon$, are defined by acting with creation operators on the vacuum:
\begin{equation}
| \epsilon \rangle = \prod_{v \in V} \phi^\ast(v)^{(\epsilon,v)}  | 0 \rangle.
\end{equation}
This element depends on the ordering of vertices in the product, but different orderings give rise to states differing only by a factor $\pm 1$. To~well-define vectors~$| \epsilon \rangle$, fix~any total order on $V$ once and for all. The set of all vectors $| \epsilon \rangle$ is an~orthonormal basis of~$\mathcal F$.

Let us define the grading element of $\mathcal{A}$:
\begin{equation}
    \gamma=\prod\limits_{v\in V}(1-2\phi^\ast(v)\phi(v)).
\end{equation}
It satisfies $\gamma=\gamma^\ast = \gamma^{-1}$. For each $\alpha\in\mathbb{Z}_2$ we define
\begin{subequations}
\begin{gather}
\mathcal F_{\alpha} = \{ \psi \in \mathcal F | \ \gamma \psi = (-1)^{\alpha} \psi \}, \\
\mathcal A_{\alpha} =\{ T \in \mathcal A | \ \gamma T = (-1)^{\alpha} T \gamma \}.
\end{gather}
\end{subequations}
$\mathcal A_0$ is a subalgebra of $\mathcal A$. Its action on $\mathcal F$ has two nontrivial invariant subspaces: $\mathcal F_0$ and $\mathcal F_1$, which are both of dimension $2^{|V|-1}$. It follows from the Artin-Weddeburn theory \cite{aw} that the algebra $\mathcal A_0$ is semisimple, with two simple factors $\mathcal A_{\alpha \alpha} = \mathrm{End}_{\mathbb C}( \mathcal F_{\alpha})$, $\alpha \in \mathbb Z_2$. This means that every finite-dimensional representation $V$ of $\mathcal A_0$ is isomorphic to $\bigoplus \limits_{\alpha \in \mathbb Z_2} \mathcal F_{\alpha}^{\oplus [V: \mathcal F_{\alpha}]}$, where multiplicity $[V : \mathcal F_{\alpha}]$ is given by the formula
\begin{equation}
[V : \mathcal F_{\alpha}] = \frac{1}{\dim_{\mathbb C}(\mathcal F_{\alpha})} \mathrm{tr}_V \left( \frac{1+ (-1)^{\alpha} \gamma}{2} \right).
\end{equation}

The even subalgebra $\mathcal{A}_0$ is of our main interest here. It is easy to see that it is generated by elements $\{ \gamma(v) \}_{v \in V}$ and $\{ \mathfrak s (e) \}_{e \in E_{\mathrm{or}}}$, defined by
\begin{equation}
\gamma(v) = 1 - 2 \phi^\ast(v) \phi(v), \qquad \mathfrak s(e) =  X(s(e)) X(t(e)) .
\end{equation}
We refer to $\gamma(v)$ and $\mathfrak s(e)$ as fermionic parity and kinetic operators, respectively. We will now give a complete set of relations satisfied by our chosen generators\footnote{More precisely, $\mathcal A_0$ is isomorphic to a quotient of the free algebra on letters $\gamma(v)$, $\mathfrak{s}(e)$ by some two-sided ideal $\mathcal I$. We will describe a set of generators of $\mathcal I$.}. Firstly,
\begin{subequations}
\begin{gather}
\gamma(v)=\gamma(v)^\ast = \gamma(v)^{-1}, \qquad \gamma(v) \gamma(v') = \gamma(v') \gamma(v), \\
-\mathfrak s(e)=\mathfrak s(\overline e)=\mathfrak s(e)^\ast= \mathfrak s(e)^{-1}, \qquad \mathfrak s(e) \mathfrak s(e') = (-1)^{(\partial e, \partial e')} \mathfrak s(e') \mathfrak s(e) \label{eq:S_braiding}, \\
\gamma(v) \mathfrak s(e) = (-1)^{(\partial e, v)} \mathfrak s(e) \gamma(v).
\end{gather}
\label{eq:A0_relations}
\end{subequations}
The final relation in $\mathcal A_0$ may be formulated as follows: if $\ell=(e_1,...,e_n)$ is a circuit, then 
\begin{equation}
\mathfrak s(e_1)\cdot ... \cdot \mathfrak s(e_n) = 1.
\label{eq:S_loop_relation}
\end{equation}
Not all of these relations are independent. Indeed, suppose that some algebra $\mathcal B$ contains elements $\gamma(v)$ and $\mathfrak s(e)$ satisfying \eqref{eq:A0_relations} and such that \eqref{eq:S_loop_relation} holds for some circuits $\{ \ell_i \}_{i=1}^s$ such that $[ \ell_i ]$ generate $Z_1$. Then for any circuit $\ell=(e_1,...,e_n)$ there exist coefficients $c_i$ such that $[ \ell ] = \sum\limits_{i=1}^s c_i [ \ell_i ]$. Using relations~\eqref{eq:A0_relations} and \eqref{eq:S_loop_relation} for $\ell_i$ we obtain $\mathfrak s(e_1)\cdot...\cdot \mathfrak{s}(e_n)  = \pm 1$. The same calculation can be repeated in $\mathcal A_0$, so the sign on right hand side has to be $+1$, because \eqref{eq:S_loop_relation} holds for all circuits in this case. Hence \eqref{eq:S_loop_relation} is satisfied in $\mathcal B$ for all~circuits~$\ell$.
 
In the rest of this section we will show that there are no other relations, i.e.\ that \eqref{eq:A0_relations} and \eqref{eq:S_loop_relation} generate all relations in $\mathcal A_0$. It will be convenient to consider operators
\begin{subequations}
\begin{gather}
\gamma(\epsilon) = \prod_{v \in V} \gamma(v)^{(\epsilon,v)}, \qquad \mathrm{for} \ \epsilon \in C_0, \\
\mathfrak s(\tau) = \prod_{e \in E} \mathfrak s(e)^{(e, \tau)}, \qquad \mathrm{for} \ \tau \in C_1.
\end{gather}
\end{subequations}
The sign of $\mathfrak s(\tau)$ depends on a choice of orientation for each $e \in E$ and an ordering of $E$, which we fix for the purpose of the proof. These operators satisfy $\gamma(\epsilon) | \epsilon' \rangle = (-1)^{(\epsilon, \epsilon')} | \epsilon' \rangle$ and $\mathfrak s(\tau) | \epsilon \rangle = (-1)^{\chi(\tau, \epsilon)} | \epsilon + \partial \tau \rangle$ for some function $\chi : C_1 \times C_0 \to \mathbb Z_2$, which depends on the arbitrary choices made.

Using relations \eqref{eq:A0_relations} only, any monomial in the generators $\mathfrak s(e)$ and $\gamma(v)$ may be rewritten (perhaps up to a sign) as a product $\gamma(\epsilon) \mathfrak s(\tau)$ for some $\epsilon \in C_0$ and $\tau \in C_1$. 

Now let $r$ be a section of $\partial : C_1 \to B_0$, i.e. a linear map $B_0 \to C_1$ such that $\partial r = 1_{B_0}$. Notice that such $r$ is guaranteed to exist, because $\partial$ is a linear map between vector spaces with image $B_0$. However, it is by no means unique.

For any $\tau \in C_1$ let $z(\tau) = \tau - r \partial \tau  \in C_1$. Then we have $\tau = r \partial \tau + z (\tau)$ and $\partial z (\tau) = 0$, so~$\mathfrak s(\tau)$ coincides with $\mathfrak s(r \partial \tau)$, possibly up to a sign. This means that, up to a sign, monomial $\gamma(\epsilon) \mathfrak s(\tau)$ depends on $\tau$ only through $\partial \tau$.

Using relations described so far, any relation in $\mathcal A_0$ may be reduced to
\begin{equation}
\sum_{\epsilon \in C_0} \sum_{\epsilon' \in B_0} c_{\epsilon, \epsilon'} \gamma(\epsilon) \mathfrak s(r\epsilon') =0,
\end{equation}
where $c_{\epsilon, \epsilon'}$ are complex coefficients. 

Acting with the operator on the left hand side on the vector $| \epsilon'' \rangle$ we obtain
\begin{equation}
\sum_{\epsilon \in C_0} \sum_{\epsilon' \in B_0} c_{\epsilon, \epsilon'}  (-1)^{(\epsilon, \epsilon' + \epsilon'')} (-1)^{\chi(r \epsilon', \epsilon'')} |\epsilon'' + \epsilon' \rangle =0.
\end{equation}

Since the set $\{ | \epsilon'' + \epsilon' \rangle \}_{\epsilon' \in B_0}$ is linearly independent in $\mathcal F$, each term of the summation over $\epsilon'$ vanishes separately. Therefore we have
\begin{equation}
\sum_{\epsilon \in C_0} c_{\epsilon, \epsilon'} (-1)^{(\epsilon, \epsilon' + \epsilon'')}=0.
\end{equation}

Now let $\epsilon_1= \epsilon' + \epsilon''$, take any $ \epsilon_2 \in C_0$ and multiply this equation by $(-1)^{(\epsilon_1, \epsilon_2)}$. Summing over all $\epsilon_1$ and using the identity $\sum\limits_{\epsilon_1 \in C_0} (-1)^{(\epsilon_1, \epsilon+ \epsilon_2)}= 2^{|V|} \delta_{\epsilon,\epsilon_2}$ we get
\begin{equation}
c_{\epsilon_2, \epsilon'} =0.
\end{equation}
Since $\epsilon_2$ and $\epsilon'$ were arbitrary, all coefficients $c$ vanish. We have shown that any relation in $\mathcal A_0$ follows already from \eqref{eq:A0_relations} and \eqref{eq:S_loop_relation}, which completes the proof.

\section{$\Gamma$ model \label{sec:gamma_model}}

We will now construct a bosonic model equivalent to the fermionic one discussed in the previous section. Relations (\ref{eq:A0_relations}) will be satisfied as operator equations, but (\ref{eq:S_loop_relation}) will be imposed as a constraint on physical states. Due to the presence of $\Gamma$ matrices in its formulation, we will refer to it as the $\Gamma$ model \cite{W}. Generators of the algebra $\mathcal A_0$ will be constructed as simple, local expressions in fields of the $\Gamma$ model. Afterwards, we propose a~correspondence between the $\Gamma$ model and a certain $\mathbb Z_2$ gauge theory. The section is closed with a discussion of the $\Gamma$ model and its constraints in case of toroidal geometries.

\subsection{Definition of the model} \label{sec:gamma_def}

In this section we will assume that the graph $\mathfrak G$ is such that every vertex has even degree. To~a~vertex $v$ we associate the Clifford algebra with generators $\{ \Gamma_\ast(v) \} \cup \{ \Gamma(v,e) \}_{e \in \mathrm{St}(v)}$. Each generator squares to identity and anticommutes with every other generator located on the same vertex, but generators on different vertices commute. Clifford algebras associated to distinct vertices may be non-isomorphic, because we do not assume that all $v \in V$ have the same degree. Secondly, we construct an irreducible representation of the algebra associated to each vertex. There is some arbitrariness here, because there exist two non-isomorphic simple modules, corresponding to two possible values of $\Gamma_\ast(v) \prod\limits_{e \in \mathrm{St}(v)} \Gamma(v,e)$. For now we make some choice for every vertex. We will discuss its significance in subsection \ref{sec:rep_ambig}. Hilbert space $\mathcal H$ of the $\Gamma$ model is defined as the tensor product of Hilbert spaces associated to individual vertices. Thus operators on distinct vertices commute. In this sense $\Gamma$ model is bosonic.

Kinetic operators of the $\Gamma$ model are defined in the following way. For every edge $e$ we choose an orientation and put
\begin{equation}
    S(e) = - i \Gamma(s(e),e) \Gamma(t(e),e).
\end{equation}
For the opposite orientation we define $S(\overline e) := - S(e)$.

A simple calculation shows that the map
\begin{equation}
 \gamma(v) \mapsto \Gamma_\ast(v), \qquad \mathfrak s(e) \mapsto S(e)
\label{eq:bosonization_map}
\end{equation}
is compatible with (\ref{eq:A0_relations}). However (\ref{eq:S_loop_relation}) does not hold as an operator relation. Nevertheless, if $\ell = (e_1,...,e_n)$ is a circuit, then $S(\ell) := S(e_1)\cdot...\cdot S(e_n)$ is unitary, squares to identity and commutes with all $\Gamma_\ast(v)$ and $S(e)$. Therefore the subspace $\mathcal H_0 \subseteq \mathcal H$ of all vectors $\psi$ satisfying the constraint
\begin{equation}
S (\ell) \psi = \psi \qquad \text{for every circuit }  \ell
\label{eq:phys_states}
\end{equation}
is a representation of the algebra $\mathcal A_0$.

We claim that $\mathcal H_0$ is isomorphic (as a representation of $\mathcal A_0$) to a half of the Fock space, i.e. $\mathcal H_0 \cong \mathcal F_{\alpha}$ for some $\alpha$. Remainder of this subsection is devoted to the proof of this fact.

Let $\ell = \left( e_1,...,e_{|E|} \right)$ be an Eulerian circuit. Then $S(\ell) = (-1)^{\alpha} \prod\limits_{v \in V} \Gamma_\ast(v)$ for some $\alpha$. Therefore acting with $S(\ell)$ on $\psi_0 \in \mathcal H_0$ we obtain
\begin{equation}
\left( \prod_{v \in V} \Gamma_{\ast}(v) \right) \psi_0 = (-1)^{\alpha} \psi_0.
\label{eq:phys_states_parity}
\end{equation}
This means that $\mathcal H_0$ is a direct sum of some number of copies of $\mathcal F_{\alpha}$. To show that the multiplicity is equal to one it is sufficient to demonstrate that $\dim (\mathcal H_0) = 2^{|V|-1}$. For this purpose let us first note that
\begin{equation}
\dim(\mathcal H) = \prod\limits_{v \in V} 2^{\frac{\deg(v)}{2}} = 2^{|E|}.
\end{equation}
Secondly, for every $[A] \in Z_1^\ast$ let $\mathcal H_{[A]}$ be the set of vectors $\psi$ such that
\begin{equation}
S(\ell) \psi = (-1)^{([A],[\ell])} \psi \qquad \text{for every circuit } \ell.
\label{eq:gauged_phys_states}
\end{equation}
We have a decomposition $\mathcal H = \bigoplus\limits_{[A] \in Z_1^\ast} \mathcal H_{[A]}$. By the formula \eqref{eq:Z1_dim} there are $2^{|E|-|V|+1}$ summands, so the problem is reduced to checking that each $\mathcal H_{[A]}$ has the same dimension. This is achieved by considering the unitary operators
\begin{equation}
O(\tau) = \prod\limits_{e \in E} \Gamma(s(e),e)^{(\tau ,e )} \qquad \mathrm{for} \ \tau \in C^1,
\label{eq:O_operators}
\end{equation}
in which we choose orientation of each $e \in E$. Simple calculation shows that they satisfy
\begin{equation}
O(\tau) S(\ell) = (-1)^{([\tau], [ \ell ])} S(\ell) O(\tau) \qquad \text{if } \ell \text{ is a circuit}.
\end{equation}
This implies that $O(\tau) \mathcal H_{[A]} \subseteq \mathcal H_{[A + \tau]}$, from which the result follows.

\subsection{Choice of a representation \label{sec:rep_ambig}}

Recall that in the construction of our model it was necessary to choose a representation of the Clifford algebra at every vertex $v$. This is equivalent to specifying a relation between the action of $\Gamma_{\ast}(v)$ and $\prod\limits_{e \in \mathrm{St}(v)} \Gamma(v,e)$. The two operators are proportional in every irreducible representation, but there are two possible values of the proportionality factor. One can resolve the ambiguity as follows. Let us choose some ordering of the set $\mathrm{St}(v)$. Then we may denote its elements as $e_1,...,e_{2n}$ with $n = \frac{\deg(v)}{2}$. Having done that we put
\begin{equation}
\Gamma_\ast(v) := i^{n} \Gamma(v,e_1) \cdot ... \cdot \Gamma(v,e_{2n}).
\end{equation}
This is a consistent definition - element $\Gamma_{\ast}(v)$ anticommutes with all $\Gamma(v,e)$ and squares to $1$. It is invariant with respect to even permutations of the indexing set $\{ 1, ..., n \}$, but it changes sign under any odd permutation. 

We see that our model is completely specified once we choose an ordering (modulo even permutations) of the set $\mathrm{St}(v)$ for each vertex $v$. We are not aware of a natural way to make this choice, save for the case of some very symmetric geometries. Thus it is crucial to understand its consequences. Any other construction of $\Gamma_{\ast}$ is related to the chosen one by
\begin{equation}
\Gamma_{\ast}'(v) = (-1)^{(\eta,v)} \Gamma_{\ast}(v)
\label{eq:eta_def}
\end{equation}
with some $\eta \in C_0$. Thus the space of distinct choices is affine over $C_0$. It does not seem to have a distinguished origin. 

Now let us consider the unitary operators
\begin{equation}
    T(\theta) = \left[ \prod_{v \in V} \Gamma_{\ast}(v)^{(\partial \theta, v)} \right] \cdot \left[ \prod_{e \in E} S(e)^{(\theta,e)} \right] \qquad \text{for } \theta \in C_1,
\end{equation}
whose signs depend on a choice of orientations of edges and an ordering of $E$. They commute with all $S(e)$ and satisfy
\begin{equation}
T(\theta) \Gamma_{\ast}(v) T(\theta)^{-1} = (-1)^{(\partial \theta, v)} \Gamma_{\ast}(v).
\end{equation}
This establishes that constructions of our model related by (\ref{eq:eta_def}) are unitarily equivalent if $\eta = \partial \theta$. Thus they describe the same physics for any choice of hamiltonian built of fermionic parity and kinetic operators. Identifying equivalent models we see that the set of distinct versions of the $\Gamma$ model is affine over the homology group $C_0 / B_0 \cong \mathbb Z_2$, or in simpler words - it has two elements. They correspond to two possible values of $\alpha$ in (\ref{eq:phys_states_parity}). Indeed, redefinition (\ref{eq:eta_def}) with $\eta$ representing a nonzero homology class (i.e.~a sum of an odd number of vertices) changes the sign of the operator $\prod\limits_{v \in V} \Gamma_\ast(v)$ while keeping the form of constraints (\ref{eq:phys_states}) invariant.

The discussion above may be phrased in the language of higher symmetries \cite{gksw} as follows: construction of our model has a sort of gauge freedom, with gauge transformations parametrized by $1$-chains. If the graph $\mathfrak G$ is the one-skeleton of a closed $d$-dimensional manifold\footnote{$X$ does not have to be orientable, because we need only Poincar\'{e} duality over $\mathbb Z_2$.} $X$, there is a Poincar\'{e}-duality between $1$-chains and $(d-1)$-cochains. In this sense $\Gamma$ model has a~$(d-1)$-form $\mathbb Z_2$ gauge invariance. We may identify $\prod\limits_{v \in V} \Gamma_{\ast}(v)$ as the unique nontrivial gauge-invariant $d$-holonomy operator. Choice of a particular representation involves fixing the gauge as well as the value of this operator.

We will now describe how to construct data needed to completely determine the $\Gamma$ model corresponding to a prescribed value of $\alpha$. It suffices to do this for $\alpha=1$, the other case being obtained by a transformation (\ref{eq:eta_def}) with any $\eta \not \in B_0$. Let $\ell = \left( e_1,...,e_{|E|} \right)$ be an Eulerian circuit. For every $v \in V$ there are exactly $n:=\frac{\deg(v)}{2}$ indices $1 \leq j_1 < j_2 < ... <j_n \leq |E|$ such that $s(e_{j_i}) = v$. We define an ordering on $\mathrm{St}(v)$ by
\begin{equation}
e_{j_1-1} < e_{j_1} < e_{j_2-1} < e_{j_2} < ... < e_{j_n-1} < e_{j_n},
\label{eq:Eulerian_Ordering}
\end{equation}
where $e_0:=e_{|E|}$. It is easy to check that then $S(\ell) = - \Gamma_\ast(v)$, so $\alpha=1$. In particular, distinct choices of the Eulerian circuit $\ell$ give rise to orderings which are equivalent in the sense described in the previous paragraph.

\subsection{Modified constraints and $\mathbb Z_2$ gauge fields} \label{sec:modified_constraints}

Consider coupling fermions to an external lattice $\mathbb Z_2$ gauge field. The gauge field is a~cochain $A \in C^1$ subject to gauge transformations $A \mapsto A + \delta \theta$ with $\theta \in C^0$. Thus gauge orbits are parametrized by equivalence classes $[A] \in C^1/B^1 = Z_1^\ast$. The minimal coupling rule asserts that each occurence of $\mathfrak s(e)$ in the fermionic hamiltonian should be replaced by $\mathfrak s_A(e) := \mathfrak s(e) \cdot (-1)^{(A,e)}$. These operators satisfy the same relations as the original $\mathfrak s(e)$ except of (\ref{eq:S_loop_relation}), which is replaced by
\begin{equation}
\mathfrak s_A(e_1) \cdot ... \cdot \mathfrak s_A(e_n) = (-1)^{([A],[\ell])} \qquad \text{for every circuit } \ell = (e_1,...,e_n).
\label{eq:gauged_loop_relation}
\end{equation}
Now consider the bosonization map
\begin{equation}
    \gamma(v) \mapsto \Gamma_\ast(v), \qquad \mathfrak s_A(e) \mapsto S(e).
    \label{eq:gauged_bosonization_map}
\end{equation}
In order for this prescription to be compatible with the relation (\ref{eq:gauged_loop_relation}) it is necessary to restrict attention to the subspace $\mathcal H_{[A]} \subseteq \mathcal H$ of vectors $\psi$ satisfying the constraint (\ref{eq:gauged_phys_states}). Notice that the form of this condition is gauge-independent, because $(A, [\ell])$ depends only on the gauge orbit $[A]$ of $A$ for every circuit $\ell$. On the other hand, the form of the bosonization map (\ref{eq:gauged_bosonization_map}) does depend on the choice of gauge.

We conclude that in order to couple fermions to a $\mathbb Z_2$ gauge field it is sufficient to change the form of constraint to (\ref{eq:gauged_phys_states}), without changing the form of hamiltonian expressed in terms of $\gamma(v)$ and $S(e)$ operators. It remains to describe the structure of the $\mathcal A_0$-module $\mathcal H_{[A]}$. We pick an Eulerian circuit $\ell$ and an element $\psi \in \mathcal H_{[A]}$. Then
\begin{equation}
    (-1)^{\alpha} \left( \prod\limits_{v \in V} \Gamma_\ast(v) \right) \psi = S(\ell) \psi = (-1)^{([A],\zeta)} \psi,
\end{equation}
where we used the fact that $[\ell] = \zeta$ for any Eulerian circuit $\ell$.

We conclude that $\mathcal H_{[A]}$ is isomorphic to a direct sum of some number of copies of $\mathcal F_{\alpha + ([A],\zeta)}$. Since $\dim (\mathcal H_{[A]}) = \dim (\mathcal H_0)$, the multiplicity is equal to one. 

We have shown that the full Hilbert space of the $\Gamma$ model decomposes as a direct sum of subspaces describing fermions coupled to all possible external $\mathbb Z_2$ gauge fields. Interestingly, the allowed value of fermionic parity depends on the "magnetic" observable $([A], \zeta)$.

One could ask whether it is possible to promote the gauge field to a dynamical degree of freedom. In order to write down a kinetic term for the $A$ field it would be necessary to invoke "electric" operators which connect subspaces corresponding to different values of the gauge field. Before answering to what extent such operators exist in our model, we briefly review the construction of the conventional $\mathbb Z_2$ gauge theory \cite{wegner,kogut} coupled to fermions. 

The Hilbert space is defined initially as the tensor product of the fermionic Hilbert space and the Hilbert space for gauge fields. The latter has an orthonormal basis $\{| A \rangle\}$ with $A$ running over all elements of $C^1$. Magnetic operators $U(\tau)$ are parametrized by chains $\tau \in C_1$. They act on basis states according to the formula $U(\tau) | A \rangle = (-1)^{(A, \tau)} | A \rangle$. Electric operators $W(\omega)$ are parametrized by $\omega \in C^1$ and defined by $W(\omega) | A \rangle = | A + \omega \rangle$. Thus one has braiding relations $U(\tau) W(\omega) = (-1)^{(\omega, \tau)} W(\omega) U(\tau)$.

In the next step one introduces Gauss' operators $G(\theta) = \gamma(\theta)W(\delta \theta)$ for $\theta \in C_0$. They implement $\mathbb Z_2$ gauge transformations. Only gauge-invariant states ($G(\theta) \psi = \psi$) are regarded as physical. This defines the true Hilbert space of the theory. Taking $\theta = \sum \limits_{v \in V} v$ one finds that all physical states are eigenvectors of $\gamma$ to eigenvalue one, so there are no states with odd number of fermions. 

The algebra of gauge-invariant operators ($G(\theta) O G(\theta)^{-1}=O$) is generated by dressed kinetic operators $\mathfrak s_g(e)=\mathfrak s(e) \cdot U(e)$ and electric operators $W(e)$. There are magnetic observables $U(\tau)$ for $\partial \tau =0$, but these may be expressed in terms of kinetic operators. Indeed, for $\ell$ being a circuit
\begin{equation}
 U([\ell]) =    \mathfrak s_g(\ell).
\end{equation}
Similarly the charge operators may be expressed\footnote{We regard gauge-invariant operators as acting on the physical Hilbert space only, so identities which follow from the Gauss' law are written as operator relations.} in terms of electric operators:
\begin{equation}
\gamma(v) = W(\delta v).
\end{equation}

The only independent relations between our chosen generators are (\ref{eq:S_braiding}) with $\mathfrak s$ replaced by $\mathfrak s_g$, the following properties of $W$:
\begin{equation}
W(\omega)= W(\omega)^\ast = W(\omega)^{-1}, \qquad W(\omega_1 + \omega_2) = W(\omega_1) W(\omega_2),
\label{eq:V_relations}
\end{equation}
and braiding relations between kinetic and electric operators
\begin{equation}
\mathfrak s_g(e) W(\omega) = (-1)^{(\omega,e)} W(\omega) \mathfrak s_g(e).
\label{eq:hop_electric_braid}
\end{equation}

Now we return to the $\Gamma$ model considered without any constraints on physical states. We ask if the algebra of gauge-invariant operators of $\mathbb Z_2$ gauge theory may be represented on its Hilbert space. We would like to map $\mathfrak s_g(e)$ to $S(e)$ and $\gamma(v)$ to $\Gamma_\ast(v)$. This is consistent with local relations in gauge theory, but it is inconsistent with the global relation $\prod \limits_{v \in V} \gamma(v) =1$, since we have instead $\prod \limits_{v \in V} \Gamma_\ast(v) = (-1)^{\alpha} S(\ell)$ for an Eulerian circuit $\ell$. On~the gauge theory side the problematic relation is a consequence of the Gauss' law, so we would like to interpret the $\Gamma$ model as a gauge theory with deformed Gauss' law. Such deformation has the consequence that it is not possible to represent operators $W(e)$ in a~way compatible with $W(e) W(e')=W(e')W(e)$ and braiding relations (\ref{eq:hop_electric_braid}), because these operators would have to anticommute with the c-number $\Gamma_\ast(v) S(\ell)=(-1)^{\alpha}$, which is absurd\footnote{One way to avoid this conclusion is to consider the direct sum of Hilbert spaces of two versions of the $\Gamma$ model corresponding to two values of $\alpha$. Then $(-1)^{\alpha}$ is promoted to an operator with eigenvalues $\pm 1$, so it is possible to introduce operators which anticommute with it. Such construction was considered in \cite{szczerba}, but this is not what we would like to do here.}. This argument does not concern operators $W(\omega)$ with $\omega$ orthogonal to $\zeta$, i.e. those $\omega$ which are sums of even numbers of edges. To construct a convenient basis of $C_1^{\mathrm{even}}$, the orthogonal complement of $\zeta$, let $\ell = (e_1,...,e_{|E|})$ be an Eulerian circuit. Put $\epsilon_i=e_{i-1} + e_{i} \in C_1$ for $2 \leq i \leq |E|$. Then $\epsilon_i$ form a basis of $C_1^{\mathrm{even}}$ and have the convenient property that each $\epsilon_i$ is a sum of two edges which meet at the vertex $v_i:=s(e_i)$. Since each edge $e \in E$ is equal to $e_i$ for exactly one $i$, this construction defines a partition of each set $\mathrm{St}(v)$ into a disjoint union of $\frac{\mathrm{deg}(v)}{2}$ pairs of the form $e_{i-1}, e_{i}$ (where $e_{0} := e_{|E|}$) with $1 \leq i \leq |E|$ such that $v=v_i$. Now define
\begin{equation}
\mathcal W(\epsilon_i) = (-1)^{\kappa_i} \cdot i \Gamma(v_i,e_{i-1}) \Gamma(v_i,e_{i}) \qquad \mathrm{for} \ 2 \leq i \leq |E|,
\end{equation}
where $\kappa_i \in \mathbb Z_2$ is not yet specified. Operators $\mathcal W(\epsilon_i)$ are our candidates for representatives of $W(\epsilon_i)$. We have $\mathcal W(\epsilon_i) \mathcal W(\epsilon_j) = \mathcal W(\epsilon_j) \mathcal W(\epsilon_i)$ and $\mathcal W(\epsilon_i)^2=1$, so we may well-define $\mathcal W(\omega)$ for any $\omega \in C_1^{\mathrm{even}}$ by demanding that $\mathcal W(\omega_1 + \omega_2) = \mathcal W(\omega_1) \mathcal W(\omega_2)$. For example
\begin{equation}
\mathcal{W}(e_1 + e_n) = \prod_{i=2}^{|E|} \mathcal W(\epsilon_i),
\end{equation}
since $e_1 + e_n = \sum\limits_{i=2}^{|E|} \epsilon_i$. With this definition relations (\ref{eq:V_relations}) and (\ref{eq:hop_electric_braid}) are satisfied. Furthermore, we can choose $\kappa_i$ in such a way that $\mathcal W(\delta v)= \Gamma_\ast(v)$ is satisfied for every vertex other than $v_1:= s(e_1)=t(e_n)$. For example if elements $\Gamma_*(v)$ are constructed as in the discussion surrounding equation (\ref{eq:Eulerian_Ordering}), one may take all $\kappa_i=0$. In any case we have
\begin{equation}
\mathcal W(\delta v_1)  =  \mathcal W \left( \sum_{v \neq v_1} \delta v \right) = \prod_{v \neq v_1} \Gamma_*(v)  = (-1)^{\alpha} S(\ell) \cdot \Gamma_*(v_1).
\end{equation}
This means that for the single vertex $v_1$ the Gauss' law is modified by the factor $(-1)^{\alpha} S(\ell)$.

We are now ready to define the gauge theory corresponding to the $\Gamma$ model with no constraints imposed. Elementary fermionic operators as well as $U$ and $W$ operators are constructed as in the conventional gauge theory. The only modification is in the definition of the Gauss' operators, which are taken to be
\begin{equation}
G(v) = 
\begin{cases}
\gamma(v) W(\delta v) & \text{for } v \neq v_1, \\
(-1)^{\alpha} \gamma(v) U(\zeta) W(\delta v) & \text{for } v = v_1.
\end{cases}
\label{eq:gauss_zeta}
\end{equation}
This has the consequence that also the algebra of gauge invariant operators is modified. We study properties of this gauge theory and its generalizations in section \ref{sec:deformed_gauss}. Here we summarize those results obtained there which are directly relevant for the correspondence with the $\Gamma$ model:
\begin{itemize}
    \item An isomorphism between the algebra of gauge-invariant operators in gauge theory and $\mathrm{End}(\mathcal H)$ is constructed. Operators constructed of even numbers of fermions and Wilson lines are mapped to local operators in the $\Gamma$ model, but electric operators are represented in a way which is non-local and depends on the choice of an Eulerian circuit. Similarly, there exist non-local operators in gauge theory corresponding to $\Gamma(v,e)$ from the $\Gamma$ model.
    \item The definition of Gauss' operators suggests that there is an inherent non-locality and lack of symmetry between distinct vertices in the proposed gauge theory. We~demonstrate that under certain assumptions about the underlying geometry these pathologies can be healed by a canonical transformation.
    \item The Gauss' law, which is imposed as a constraint in the gauge theory picture, holds identically in the $\Gamma$ model. Therefore all states and all operators in the $\Gamma$ model are gauge invariant.
    \item Relation between the total number of fermions mod $2$ and the value of $[A]$ satisfied in the $\Gamma$ model is a consequence of the Gauss' law on the gauge theory side.
\end{itemize}

It is interesting to interpret the algebra of $\{ \Gamma(v,e) \}$, the elementary fields of the $\Gamma$ model, in terms of quantum numbers defined in gauge-theoretical language. To this end we inspect the braiding relations
\begin{subequations}
\begin{gather}
\Gamma_*(v')\Gamma(v,e) = (-1)^{(v,v')} \Gamma(v,e) \Gamma_*(v'), \\
S(\ell) \Gamma(v,e) = (-1)^{([\ell],e)} \Gamma(v,e) S(\ell) \qquad \text{if } \ell \text{ is a circuit}.
\end{gather}
\end{subequations}
The first relation asserts that $\Gamma(v,e)$ flips the value of fermionic parity at the vertex $v$, i.e.\ it creates or annihilates a fermion. The second one means that action of $\Gamma(v,e)$ changes the value of the holonomy along any loop which contains the edge $e$. There is no operator that creates or annihilates a single fermion without disturbing the values of holonomies or a one that acts as an electric field operator on a single edge without creating any fermions, because that would contradict the relation
\begin{equation}
\text{Total number of fermions}\pmod{2}\, =\, \alpha + ([A], \zeta).
\label{eq:fermi_parity_A_rel}
\end{equation}
The preceding discussion justifies thinking of $\Gamma(v,e)$ as a composite of a~fermion and a lump of electromagnetic field, as in the so-called flux attachment mechanism.

According to the presented picture, the role of constraints \eqref{eq:phys_states} present in our bosonization map is to get rid of the electromagnetic degrees of freedom present in the $\Gamma$ model. We~close this discussion with the remark that constraints can be divided into two classes:
\begin{enumerate}
    \item Constraints which correspond to homologically trivial loops, i.e.\ circuits $\ell$ such that the cycle $[\ell]$ belongs to $B_1$. It is sufficient to impose one such constraint for every face of the lattice. Constraints of this type are local, and hence can be implemented by introducing in the hamiltonian local terms which penalize their violation. They reduce the Hilbert space from $\mathcal H$ to the direct sum of subspaces corresponding to gauge orbits of flat gauge fields, i.e.~to $\bigoplus \limits_{[A] \in H^1} \mathcal H_{[A]}$.
    \item Constraints which correspond to loops of nonzero homology class. Once constraints of the first type are imposed, operators corresponding to distinct representatives of the same homology class become equivalent. It is sufficient to impose one such constraint for every element of some basis of $H_1$. This chooses from the set of all flat gauge fields the trivial gauge field $A=0$.
\end{enumerate}

\subsection{Example: toroidal geometries} \label{sec:toroidal}

In this subsection we construct the $\Gamma$ model on a torus with $L_1 \times ... \times L_d$ lattice sites, with each $L_i \geq 3$. In the case of $d=2$ and even $L_i$ we present a full solution of constraints \eqref{eq:phys_states}.

Lattice vertices are labeled by $d$-tuples of integers, with two $d$-tuples identified if they differ by a tuple whose $i$-th entry is a multiple of $L_i$ for each $i$. Sets of edges and faces are the obvious ones. Clearly every vertex has even degree. 

Operator $\Gamma(v,e)$ with edge $e$ in positive or negative $i$-th direction is denoted by $\Gamma_{\pm i}(v)$. Furthermore, we introduce
\begin{equation}
\Gamma_*(v) = (-1)^{(\eta, v)} \cdot i^d \prod_{i=1}^d \Gamma_i(v) \Gamma_{-i}(v)
\end{equation}
where $\eta$ is a $0$-chain. With this convention
\begin{equation}
    \alpha = \sum_{v \in V} (\eta, v) + \sum_{i=1}^d \prod_{j \neq i} L_j,
\end{equation}
as can be easily evaluated by computing the product $\prod \limits_{\ell} S(\ell)$ with $\ell$ running through the set of all straight lines winding once around the torus.

\begin{figure}[h!]
  \includegraphics[scale=0.6,trim={0 19cm 0cm 4cm}] {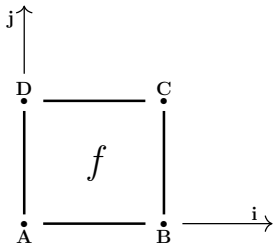}
\caption{Labels of vertices for a face $f$ lying in the plane spanned by directions $i,j$.}
\label{Figure1}
\end{figure}

Let $f$ be a~face lying in the plane spanned by directions $1 \leq i < j \leq d$, with vertices $A, B , C, D$ ordered counterclockwise, starting from the south-west corner (see figure \ref{Figure1}). 

The constraint \eqref{eq:phys_states} for the circuit around the boundary of $f$ is of the form
\begin{subequations}
\begin{gather}
\mathcal P(f) | \mathrm{phys} \rangle = | \mathrm{phys} \rangle, \label{eq:plaquette_constraint} \\
\mathcal P(f) = - \Gamma_{i,j}(A) \Gamma_{j,-i}(B) \Gamma_{-i,-j}(C) \Gamma_{ -j,i}(D) ,
\end{gather}
\end{subequations}
where $\Gamma_{k,l}(v):=\Gamma_k(v) \Gamma_l(v)$. We note the mnemonic rule that in the above, indices $\pm i, \pm j$ labeling gamma matrices are arranged in a cycle.

\begin{figure}[H]
\centering
\includegraphics[scale=1.5] {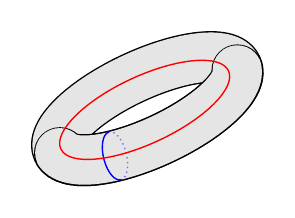}
\caption{Two loops wrapping the $2$-dimensional torus.}
\label{fig:torus1}
\end{figure}

The~only other constraints correspond to $d$ independent loops wrapping around the whole torus (see figure \ref{fig:torus1}). 
They take the form
\begin{subequations}
\begin{gather}
\mathcal L_j(v) | \mathrm{phys} \rangle = | \mathrm{phys} \rangle, \qquad j = 1, ..., d, \label{eq:loop_constraint}\\
\mathcal L_j(v) := - i^{L_j} \prod_{k=0}^{L_j-1} \Gamma_{j,-j}(t_j^k \cdot v),
\end{gather}
\end{subequations}
where $v \in V$ is a reference vertex and $t_i$ is the transformation of $V$ defined by
\begin{equation}
t_i \cdot (v_1,...,v_d)=(v_1,..., v_i +1, ..., v_d).
\end{equation}
We note that $\mathcal L_i$ are unitary, hermitian and commute with each other.

We now confine ourselves to the case of $d=2$ and all $L_i$ even. Consider the operators
\begin{subequations}
\begin{gather}
\Xi_1(v) =  \prod_{k=0}^{L_2-1} \Gamma_{1, (-1)^k 2}(t_2^k \cdot v), \\
\Xi_2(v) = \prod_{k=0}^{L_1-1} \Gamma_{(-1)^k 1, 2}(t_1^k \cdot v).
\end{gather}
\end{subequations}
They are unitary, hermitian and commute with all $\mathcal P(f)$, $\Gamma_\ast(v)$ and with each other. Moreover, they flip the values of corresponding $\mathcal L_j$:
\begin{equation}
\Xi_i(v) \mathcal L_j(v) = (-1)^{\delta_{i,j}} \mathcal L_j(v) \Xi_i(v).
\end{equation}
This means that pairs $\{ \mathcal L_1(v), \Xi_1(v) \}$ and $\{ \mathcal L_2(v), \Xi_2(v) \}$ generate two independent copies of the Pauli algebra. Thus solutions of plaquette constraints are organized in quadruplets, each of which contains precisely one solution of the loop constraint (\ref{eq:loop_constraint}). Given any state in such a quadruplet, the desired state satisfying (\ref{eq:loop_constraint}) may be easily obtained by acting with an appropriate element of the algebra generated by $\mathcal L_i$ and $\Xi_i$.

We remark that similar trick can be applied for other geometries, including higher dimensions, provided that the cycle $\zeta$ is a boundary. The role of $\Xi_i$ is played by electric operators $\mathcal W(\tau)$ with $\delta \tau =0$. These exist because $(\tau, \zeta)=0$ for $\tau \in Z^1$, $\zeta \in B_1$.

Having dealt with the loop constraints, we proceed to the analysis of plaquettes. It~will be convenient to divide the lattice into two complementary alternating sublattices, called even and odd. For example we may declare vertex $v =(v_1,v_2)$ to be even if $v_1 + v_2 = 0 \pmod{2}$. Parity of a face $f$ is defined as the parity of its south-west corner.

We will construct solutions of constraints which are simultaneous eigenvectors of $\Gamma_\ast(v)$ to eigenvalues $(-1)^{(\eta,v)}$. Solutions with other eigenvalues may then be obtained by acting with kinetic operators, which commute with all constraints. After this restriction, we have the relation $\Gamma_{1,-1}(v)\Gamma_{2,-2}(v)=-1$ for every vertex $v$. This can be used to simplify the plaquette constraints to the form
\begin{equation}
\mathcal P(f) = \Gamma_{1,2}(A) \Gamma_{1,2}(C) \Gamma_{1,-2}(B) \Gamma_{1,-2}(D). 
\end{equation}
Now we introduce new local operators by the formulas
\begin{equation}
\sigma_3(v) = \begin{cases}
i \Gamma_{1,2}(v) & \mathrm{for} \ v \ \mathrm{even}, \\
i \Gamma_{1,-2}(v) & \mathrm{for} \ v \ \mathrm{odd},
\end{cases} \qquad
\sigma_1(v) = \begin{cases}
-i \Gamma_{1,-2}(v) & \mathrm{for} \ v \ \mathrm{even}, \\
i \Gamma_{1,2}(v) & \mathrm{for} \ v \ \mathrm{odd}.
\end{cases}
\end{equation}

Then with the definition $\sigma_2(v) = -i \sigma_3(v) \sigma_1(v)$ we have
\begin{equation}
    \sigma_2(v) = i \Gamma_{1,-1}(v) \qquad \text{for every } v \in V.
    \label{eq:sigma2_def}
\end{equation}
One can check that for each $v$ operators $\{ \sigma_i(v) \}_{i=1}^3$ satisfy the standard relations obeyed by Pauli matrices, which justifies the chosen notation.

In terms of the new variables, plaquette operators take the form
\begin{equation}
\mathcal P(f) = \begin{cases}
\prod\limits_{v \in \{ A, B, C, D \}} \sigma_3(v) & \mathrm{for} \ f \ \mathrm{even}, \\
\prod\limits_{v \in \{ A, B, C, D \}} \sigma_1(v) & \mathrm{for} \ f \ \mathrm{odd}.
\end{cases}
\end{equation}
In this form plaquette constraints are readily recognized as equations defining ground states of the famous Kitaev's toric code \cite{kitaev}. It is well-known that there exist four solutions, corresponding to two values of $\mathcal L_1$ and $\mathcal L_2$. This is also in accord with our general finding about the $\Gamma$ model. For completeness we provide a prescription to construct these states in the next paragraph. 

\begin{figure}[h!]
\centering
\includegraphics[scale=0.5,trim={0cm 14cm 0cm 4cm}] {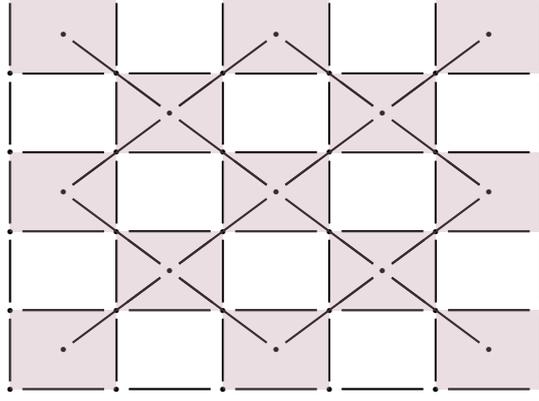}
\caption{Lattice whose vertices are the centres of even (shaded) faces of the original lattice. Its~edges and faces correspond to vertices and odd (white) faces of the original lattice, respectively.}
\label{Figure2}
\end{figure}

We work in the standard eigenbasis of $\sigma_3(v)$ operators, so our basis states are labeled by elements $\omega \in C_0$ and satisfy
\begin{subequations}
\begin{gather}
    \sigma_3(v) | \omega \rangle = (-1)^{(\omega,v)} | \omega \rangle, \\
    \sigma_1(v) | \omega \rangle = | \omega +v \rangle. \label{eq:sigma1_action}
\end{gather}
\end{subequations}
In order to have $\mathcal P(f) | \omega \rangle = | \omega \rangle$ for even faces $f$, we need to have 
\begin{equation}
(\omega, A + B + C +D) =0,
\end{equation}
where $A,B,C,D$ are the four vertices of any even face. Every such chain $\omega$ will be called admissible. Geometrically this condition means that $\omega$ may be identified with a $1$-cocycle on the lattice whose vertices are the even faces of the orignal lattice (see figure \ref{Figure2}).

Calculation analogous to the proof of \eqref{eq:Z1_dim} shows that there exist $\frac{L_1 L_2}{2}+1$ admissible chains. Now~consider the state 
\begin{equation}
| \mathrm{ref} \rangle = 2^{- \frac{L_1 L_2 +2}{4}} \sum\limits_{\omega \ \mathrm{admissible}} | \omega \rangle.
\label{eq:ref_vector}
\end{equation}
Clearly we have $\mathcal P(f) | \mathrm{ref} \rangle = | \mathrm{ref} \rangle$ for every face $f$ and $\langle \mathrm{ref} | \mathrm{ref} \rangle =1$. 

State $| \mathrm{ref} \rangle$ satisfies all plaquette constraints, but does not satisfy the loop constraints. In this paragraph we solve this difficulty. As a~fist step towards this goal, we express $\mathcal L$ and $\Xi$ operators in terms of~Pauli matrices. We take the reference vertex $v$ to be even. Then
\begin{subequations}
\begin{align}
    \mathcal L_1 (v) &= - \prod_{k=0}^{L_1 -1} \sigma_2(t_1^{k} \cdot v), \qquad
    \Xi_1(v) = (-1)^{\frac{L_2}{2}} \prod_{k=0}^{L_2-1} \sigma_3(t_2^k \cdot v), \\
   \mathcal L_2(v) &= - \prod_{k=0}^{L_2-1}  \sigma_2(t_2^k \cdot v),
 \qquad \Xi_2(v) = (-1)^{\frac{L_1}{2}} \prod_{k=0}^{L_1-1} \sigma_3(t_1^k \cdot v).
\end{align}
\end{subequations}
Using the above and the definition of $| \mathrm{ref} \rangle$ we obtain eigenvalue equations
\begin{equation}
\mathcal L_1(v) \Xi_2(v) | \mathrm{ref} \rangle = \mathcal L_2(v) \Xi_1(v) | \mathrm{ref} \rangle = - | \mathrm{ref} \rangle. 
\end{equation}
This eigensystem combined with the relations obeyed by $\mathcal L$ and $\Xi$ operators implies that projection of $| \mathrm{ref} \rangle$ onto the joint eigenspace of $\mathcal L_1$ and $\mathcal L_2$ to eigenvalue $1$ has norm $\frac{1}{2}$. To~obtain a properly normalized state, we multiply this projection by $2$: 
\begin{equation}
    | 0 \rangle = 2 \cdot  \frac{ 1 + \mathcal L_1(v)}{2} \frac{ 1 + \mathcal L_2(v) }{2} | \mathrm{ref} \rangle.
    \label{eq:zero_vector}
\end{equation}

\begin{figure}[H]
\centering
\includegraphics[scale=1.2] {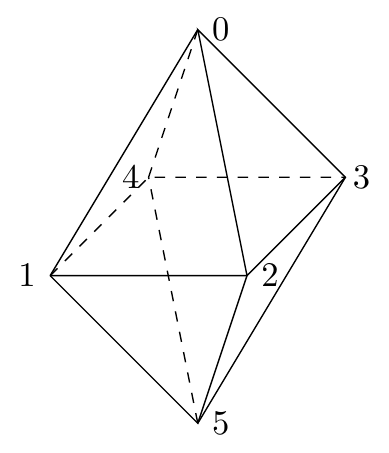}
\caption{Octahedron}
\label{fig:octahedron}
\end{figure}

We close this section with a remark that the presented method of solving constraints can be generalized to all geometries such that there exists a partition of the set of faces (say,~into "white" and "shaded" faces) such that no two faces of the same colour share an edge. Then one can construct a basis of solutions of "shaded" constraints consisting of products states, which are permuted by the action of "white" constraints. Thus the sum of all elements of this basis satisfies constraints of both types. One particularly simple decomposition of the two-sphere for which this can be carried out is the octahedron, see figure \ref{fig:octahedron}. Unfortunately, the relevant condition is never satisfied in the case of geometries of dimension higher than two. One can always obtain a solution of all constraints by acting with the projection operator $\prod \limits_f  \frac{1+ \mathcal P(f)}{2}$ on some reference state, but this does not lead to a~description as explicit as in \eqref{eq:ref_vector} and \eqref{eq:zero_vector}.

\subsection{Example: quadratic fermionic hamiltonians} \label{sec:quadratic}

Here we illustrate the bosonization procedure by applying it to hamiltonians of the form
\begin{equation}
    H=\sum\limits_{e\in E_{\mathrm{or}}}h_e \ \phi(s(e))\phi(t(e))^\ast+\sum\limits_{v\in V}\nu_v \ \phi(v)^\ast\phi(v),
\end{equation}
where $h_{\overline e}= \overline{h_e}$, while $\nu_v$ are real. This hamiltonian may be rewritten as 
\begin{equation}
    H=\sum\limits_{e\in E_{\mathrm{or}}}h_e \ \frac{1+\gamma(s(e))}{2}\mathfrak{s}(e) \frac{1+\gamma(t(e))}{2}+\sum\limits_{v\in V}\nu_v \ \frac{1-\gamma(v)}{2},
\end{equation}
from which we read off the bosonized form:
\begin{equation}
    H_{\Gamma}=\sum\limits_{e\in E_{\mathrm{or}}}h_e \ \frac{1+\Gamma_\ast(s(e))}{2}S(e)\frac{1+\Gamma_\ast(t(e))}{2}+\sum\limits_{v\in V}\nu_v \ \frac{1-\Gamma_\ast(v)}{2}.
\end{equation}
This hamiltonian commutes with $S(\ell)$ for every circuit $\ell$. Thus it has a local symmetry generated by operators $\mathcal P(f)$, which are defined as $S(\ell)$ with $[\ell]=\partial f$, and further operators labeled by loops whose classes generate the homology group $H_1$.

We will now describe the spectrum of $H_{\Gamma}$. First, consider the one-particle subspace of the fermionic system. It is governed by the $|V| \times |V|$ matrix $\{ \langle v' | H | v \rangle \}_{v,v' \in V}$. Denote its eigenvalues by $\lambda_i[h,\nu]$, $i=1,...,|V|$. The eigenvalues of $H$ are $\lambda_I[h, \nu] = \sum \limits_{i \in I} \lambda_i[h, \nu]$, indexed by subsets $I$ of $\{ 1, ..., |V| \}$. Eigenvalues of $H_{\Gamma}$ restricted to the subspace $\mathcal H_{0}$ are exactly $\lambda_I[h, \nu]$, with a restriction $|I| = \alpha \pmod{2}$. To understand the spectrum of $H_{\Gamma}$ acting on $\mathcal H_{[A]}$ with $[A] \neq 0$ notice that minimal coupling to a $\mathbb Z_2$ gauge field amounts to replacing $h_e$ by $h^A_e = h_e \cdot (-1)^{(A,e)}$. Therefore the eigenvalues of $H_{\Gamma}$ in $\mathcal H_{[A]}$ are $\lambda_I[h^A, \nu]$ with $|I| = \alpha + ([A], \zeta) \pmod{2}$. 

To enforce the plaquette constraints dynamically, consider adding to $H_{\Gamma}$ the local term
\begin{equation}
H_{c} = J \sum_{f} \frac{1- \mathcal P(f)}{2}.
\end{equation}
This leaves unchanged the eigenvalues $\lambda_I[h^A,\nu]$ for flat gauge fields $A$ and increases every other eigenvalue by at least $J$. Thus for $J$ large enough all low energy eigenstates correspond to flat gauge fields.

\section{Deformed $\mathbb Z_2$ gauge theories} \label{sec:deformed_gauss}

In this section we demonstrate that the gauge theory proposed in subsection \ref{sec:modified_constraints} is indeed equivalent to the $\Gamma$ model, even though the correspondence is local only for some operators. The proof relies on technical facts presented in the appendix \ref{canonical_heisenberg}. Afterwards we present a~certain generalization of this model, in which Gauss' operators of conventional $\mathbb Z_2$ gauge theory are modified by including phases depending on values of the holonomies. Similar mechanism is present in the Dijkgraaf-Witten theory and has been applied in the bosonization map introduced in \cite{ckr,ck,chen}. In contrast to Dijkgraaf-Witten models, here we are not restricting attention to topological gauge theories\footnote{In other words, we are working with principal bundles over $1$-skeleta which do not necessarily extend to the $2$-skeleton of the underlying space. Secondly, considered models depend on a~choice of an arbitrary $1$-cycle. We would expect only $1$-cycles dual to characteristic classes to appear in topological field theories.}. Modified Gauss' operators are classified up to (in general non-local) canonical transformations. We use this result to show how the gauge theory corresponding to the $\Gamma$ model can be formulated in a local way.

\subsection{Gauge invariant operators \label{sec:gauge_invariant_ops}}

We will now describe the algebra of gauge invariant operators for Gauss' operators of the form \eqref{eq:gauss_zeta} and explain how it is represented on the $\Gamma$ model Hilbert space. 

Operator built of $\{ X(v), Y(v) \}_{v \in V}$ will be said to be of charge $q \in C_0$ and denoted by the generic symbol $\Upsilon(q)$ if it satisfies the braiding relation
\begin{equation}
    \gamma(v)\Upsilon(q)=(-1)^{(q,v)}\Upsilon(q)\gamma(v).
\end{equation}

Every operator may be written down as a linear combination of operators of the form $\mathcal{O}=U(\tau)W(\sigma)\Upsilon(q)$. All such operators are eigenvectors of the group of gauge transformations, so the most general gauge-invariant operator is a linear combination of operators of the form $\mathcal O$ with each term separately gauge invariant. We proceed to find conditions for gauge invariance of $\mathcal O \neq 0$. Its braiding with Gauss' operators is given by
\begin{equation}
G(v)\mathcal{O}G(v)^{-1}=(-1)^{(\partial \tau, v)+(v_1,v)(\zeta,\sigma)+(q,v)}\mathcal{O},
\end{equation}
so gauge invariance of $\mathcal O$ is equivalent to the equation 
\begin{equation}
\partial \tau = q + (\zeta, \sigma) v_1.
\end{equation}
Contracting this relation with $\sum\limits_{v\in V}v$ we infer $\sum\limits_{v\in V}(q,v)=(\zeta,\sigma)$. Thus there are two possibilities: $q$ is a sum of an even or odd number of vertices. 

In the former case $(\zeta,\sigma)=0$ (so~$\sigma$ is a sum of an even number of edges) and $\partial \tau=q$. Operator $\mathcal O$ of this type is a product of 
\begin{itemize}
    \item[(a)] Wilson lines, which are allowed to terminate at charges in the usual way,
    \item[(b)] $W(\sigma)$ with $(\zeta,\sigma)=0$.
\end{itemize}
These two factors of $\mathcal O$ are separately gauge invariant. 

In the case that $q$ contains an odd number of vertices, we need $(\zeta,\sigma)=1$ and hence $\partial \tau=q+v_1$. Thus $\mathcal O$ is a~product of an~operator of the former type and $X(v_0) W(e)$ with some edge $e$.

In order to construct a set of generators convenient for comparisons with the $\Gamma$ model, choose an Eulerian circuit $\ell=(e_1,...,e_{|E|})$. We put $v_i=s(e_i)$ ($1 \leq i \leq |E|$), $\epsilon_{i}=e_{i-1}+e_i$ ($2\le i \le |E|$) and $e_0=e_{|E|}$. The algebra under consideration is generated by the set $\{\mathfrak{s}_g(e_i)\}_{i=1}^{|E|}\cup \{W(\epsilon_i)\}_{i=2}^{|E|} \cup \{ K \}$, where $K=X(v_1)W(e_0)$. Operators $U(\tau)$ for $\tau\in Z_1$ can be expressed in terms of $\{\mathfrak{s}_g(e_i)\}$, while $\gamma(v)$ is, perhaps up to a sign or a factor $U(\zeta)$, the product of some number of $W(\epsilon_i)$. The following relations are satisfied:
\begin{subequations}
\begin{gather}
-\mathfrak s_g(e_i)=\mathfrak s_g(e_i)^\ast= \mathfrak s_g(e_i)^{-1}, \qquad \mathfrak s_g(e_i) \mathfrak s_g(e_j) = (-1)^{(\partial e_i, \partial e_j)} \mathfrak s_g(e_j) \mathfrak s_g(e_i) , \\
W(\epsilon_i)=W(\epsilon_i)^\ast=W(\epsilon_i)^{-1}, \qquad W(\epsilon_i)W(\epsilon_j)=W(\epsilon_j)W(\epsilon_i),\\
\mathfrak s_g(e_i)W(\epsilon_j)=(-1)^{(e_i,\epsilon_j)}W(\epsilon_j)\mathfrak s_g(e_i),\\
K=K^\ast=K^{-1},\\
K\mathfrak{s}_g(e_i)=(-1)^{(e_i,e_0+\delta v_1)}\mathfrak s_g(e_i)K, \qquad KW(\epsilon_i)=W(\epsilon_i)K.
\end{gather}
\label{eq:A0_relations_gauged}
\end{subequations}
We have already verified that the map
\begin{equation}
    \mathfrak{s}_g(e_i)\mapsto S(e_i), \qquad W(\epsilon_i)\mapsto \mathcal{W}(e_i),
    \label{eq:gamma_map_1}
\end{equation}
defined in subsection \ref{sec:modified_constraints}, preserves all relations above not involving $K$. Thus it remains only to propose a representative of $K$ in the $\Gamma$ model. One can choose simply
\begin{equation}
    K \mapsto \Gamma(v_1,e_0),
    \label{eq:gamma_map_2}
\end{equation}
which is consistent with relations \eqref{eq:A0_relations_gauged}.

We claim that the proposed map well-defines an isomorphism between the algebra of gauge-invariant operators discussed here and the full operator algebra of the $\Gamma$ model. We~now proceed to the~proof of this fact\footnote{Consult appendix \ref{canonical_heisenberg} at this point.}. First, let us observe that relations \eqref{eq:A0_relations_gauged} are exactly as in the definition of the Heisenberg group $H_Q$ associated to a certain vector space $M$ of dimension $2n$, equipped with a quadratic form $Q$. Images of $K$ and $\{ \mathfrak s_g(e_i) \}_{i=2}^{|E|}$ in $M$ span an isotropic subspace of dimension $|E|$, so $\mathrm{Arf}(Q)=0$. Element $z \in H_Q$ acts as multiplication by $-1$, so $H_Q$ is represented faithfully. Hence all relations satisfied in the algebra of gauge-invariant operators follow already from \eqref{eq:A0_relations_gauged}. This means that equations (\ref{eq:gamma_map_1},\ref{eq:gamma_map_2}) well-define an injective homomorphism of algebras. Dimensional considerations show that this homomorphism is also surjective and that $\mathcal H$ is isomorphic to a single copy of the standard representation of $H_Q$.

Since the algebra of gauge invariant operators is isomorphic to $\mathrm{End}(\mathcal H)$, it is possible to construct an operator corresponding to $\Gamma(v,e)$ for any vertex $v$ and any $e \in \mathrm{St}(v)$. It is the product of $K$ and some number of $W(\epsilon_i)$ and $\mathfrak s_g(e_i)$, which is typically highly nonlocal.

\subsection{Classification of Gauss' operators\label{sec:classifaction}}

In the subsection \ref{sec:gauge_invariant_ops} we have considered a specific form of Gauss' operators motivated by our study of the $\Gamma$ model. In this subsection we define and classify a larger class of gauge theories. This puts previous findings in a broader context and can be applied to discuss issues with locality of our models. We are interested in gauge theories with fermionic degrees of freedom on vertices and Ising degrees of freedom $U(e)$, $W(e)$ on edges. The full Hilbert space is assumed to be endowed with a~unitary representation of the group of gauge transformations, i.e.\ for every vertex $v$ there is given a~unitary operator $G(v)$ such that $G(v)^2=1$ and $G(v)G(v') = G(v')G(v)$. Furthermore, we would like fermionic operators and $U(e)$ to transform under gauge transformations in the same way as in the conventional $\mathbb Z_2$ gauge theory, so that Wilson lines which are either closed or terminate at charges are gauge-invariant operators. This condition implies that $G(v)$ has to be of the form $\gamma(v)R(v) W(\delta v)$, where $R(v)$ is a function of operators $U(e)$ only. For~simplicity we shall assume that $G(v)$ are of particularly simple form
\begin{equation}
G(v) = (-1)^{(\mu,v)} \gamma(v) U(\mathcal T v) W(\delta v),
\end{equation}
with some $\mu\in C_0$ and $\mathcal{T}\in\mathrm{Hom}(C_0, C_1)$. Condition
$G(v)^2=1$ implies that $\mathcal{T}$ has to satisfy $(\partial \mathcal{T}v,v)=0$. Equation $G(v)G(v')=G(v')G(v)$ (for $v,v'\in V$) is equivalent to $(v,\partial\mathcal{T}v')=(v',\partial\mathcal{T}v)$. Thus $\partial \mathcal T$ is alternating, i.e.\ $(\theta, \partial \mathcal T \theta)=0$ for every $\theta \in C_0$.

Theories with Gauss' operators related by a canonical transformation of the Heisenberg group generated by $\{ U(e), W(e) \}_{e \in E}$ will be regarded as equivalent. This is a weak form of equivalence, since the allowed canonical transformations may be strongly non-local. Nevertheless it is true that equivalent theories have isomorphic algebras of gauge-invariant operators, since canonical transformations are implementable on representations of the Heisenberg group.

We wish to preserve the form of holonomy operators ($U(\tau)$ for $\partial \tau =0$), so we consider canonical transformations of the form
\begin{equation}
    U(e)\mapsto U(e), \qquad W(e)\mapsto (-1)^{(\theta, e)}U(\mathcal{S}e)W(e),
    \label{eq:allowed_canonical}
\end{equation}
for $\theta\in C_0$ and $\mathcal{S}\in\mathrm{Hom}(C_1,C_1)$. In order for this to define a canonical transformation, $\mathcal S$~must be alternating. Under a transformation of this form, $\mathcal T$ changes according to
\begin{equation}
\mathcal T \mapsto \mathcal T'= \mathcal T + \mathcal S \delta,
\end{equation}
while $\mu$ changes to some $\mu'$, which we ignore for now. It is easy to check that $\partial \mathcal S \delta$ is indeed automatically alternating if $\mathcal S$ is. Therefore the space of equivalence classes of allowed $\mathcal T$ is the quotient $\mathcal Z / \mathcal B$, where 
\begin{subequations}
\begin{gather}
 \mathcal{Z}=\{\mathcal{T}:C_0\rightarrow C_1 | \ \partial \mathcal{T} \mbox{ is alternating} \},
 \\
 \mathcal{B}=\{\mathcal{T}:C_0\rightarrow C_1  |  \mbox{ exists }\, \mathcal{S}:C_1\rightarrow C_1 \mbox{ alternating and such that } \mathcal{T}=\mathcal{S}\delta\}.
\end{gather}
\end{subequations}
Now consider the class of Gauss' operators with a fixed $\mathcal T$. They are parameterized by chains $\mu \in C_0$. However there is a residual freedom of canonical transformations with $\mathcal S = 0$ and arbitrary $\theta$. Under such transformations $\mu$ changes to $\mu' = \mu + \partial \theta$. Therefore there are two non-equivalent choices of $\mu$, corresponding to two elements of $C_0 / B_0 \cong \mathbb Z_2$.

We claim that the dimension of $\mathcal Z / \mathcal B$ is equal to $\dim (Z_1)$. For clarity we postpone the proof of this until the next paragraph. We will now establish a concrete one-to-one correspondence between pairs $(\tau, \alpha) \in Z_1 \times \mathbb Z_2$ and equivalence classes of Gauss' operators. For a given $(\tau, \alpha)$ we choose a vertex $v_1 \in V$ and define:
\begin{equation}
G(v) = 
\begin{cases}
\gamma(v) W(\delta v) & \text{for } v \neq v_1, \\
(-1)^{\alpha} \gamma(v) U(\tau) W(\delta v) & \text{for } v = v_1. 
\end{cases}
\label{eq:Gauss_nonlocal}
\end{equation}
With this definition one has
\begin{equation}
\prod_{v \in V} G(v) = (-1)^{\alpha} \gamma \cdot  U(\tau).
\end{equation}
These elements are invariant with respect to canonical transformations of the form (\ref{eq:allowed_canonical}), which demonstrates that distinct pairs $(\tau, \alpha)$ give Gauss' operators in different equivalence classes. Since the number of elements of $Z_1 \times \mathbb Z_2$ is equal to the number of equivalence classes, the one-to-one correspondence is established. There are two conclusions from this result that we would like to emphasize. Firstly, every equivalence class can be represented by $\mathcal T$ such that $\partial \mathcal T$ is not only alternating, but actually vanishes. Secondly, each equivalence class is uniquely characterized by the corresponding value of the "global" gauge transformation operator $\prod \limits_{v \in V} G(v)$, and thus by $\tau = \sum \limits_{v \in V} \mathcal T v $ and $\alpha$. If $\partial \mathcal T = 0$, one has $\alpha = \sum \limits_{v \in V} (\mu ,v)$. 

In the remainder of this subsection we calculate the dimension of $\mathcal Z / \mathcal B$. First notice that $\mathcal B$ may be identified with the quotient of the space of alternating $\mathcal S : C_1 \to C_1$ by the subspace of those $\mathcal S$ for which $\mathcal S \delta =0$. The former space has dimension $\frac{|E|(|E|-1)}{2}$. As~for the latter, any of its elements satisfies also $\partial \mathcal S=0$. Therefore it may be regarded as an alternating map $C^1/B^1 \to Z_1$. Since $C^1 / B^1 \cong Z_1^\ast$, the pertinent dimension is equal to $\frac{\dim(Z_1)(\dim(Z_1)-1)}{2}$. Hence
\begin{equation}
\dim (\mathcal B) = \frac{|E|(|E|-1)}{2} - \frac{\dim(Z_1)(\dim(Z_1)-1)}{2}.
\end{equation}
It remains to find the dimension of $\mathcal{Z}$. We consider the linear map 
\begin{equation}
    L_\partial  : \mathrm{Hom}(C_0,C_1)\ni \mathcal{T}\longmapsto \partial \mathcal{T}\in\mathrm{Hom}(C_0,B_0).
\end{equation}
Clearly $L_{\partial}$ is surjective. Secondly, $\mathrm{ker}(L_{\partial}) = \mathrm{Hom}(C_0, Z_1)$, so $\dim \ker (L_{\partial}) = \dim(Z_1) \cdot |V|$.

Next consider the space $\mathfrak{R}=\{\mathcal{R}\in\mathrm{Hom}(C_0,B_0) | \ \mathcal{R} \text{ is alternating}\}$. Choosing $V$ as a~basis of $C_0$, elements of $\mathfrak R$ are represented as symmetric $|V|\times|V|$ matrices with zeros on the diagonal and such that sum of entries in every column is $0\pmod{2}$. Simple counting\footnote{The number of free parameters in the first column is equal to $|V|-2$, since the first entry vanishes and the second one is determined in terms of the other by the requirement that the sum is even. In every subsequent column the number of free parameters decreases by one because the matrix is symmetric.} shows that $\dim(\mathfrak{R})=\frac{1}{2}(|V|-1)(|V|-2)$. Since $\mathcal{Z}=L_\partial^{-1}\mathfrak{R}$, we get 
\begin{equation}
    \dim\mathcal{Z}=\dim(\ker L_\partial) + \dim(\mathfrak{R})=|V| \dim(Z_1)+\frac{1}{2}(|V|-1)(|V|-2).
\end{equation}
Finally we use the fact that $\dim(Z_1)=|E|-|V|+1$ to simplify
\begin{equation}
\dim(\mathcal{Z}/\mathcal{B})=\dim(\mathcal{Z})-\dim(\mathcal{B})=\dim(Z_1).
\end{equation}

\subsection{Local formulations\label{sec:local}}

Gauge theories defined by Gauss' operators of the form (\ref{eq:Gauss_nonlocal}) are unsatisfactory for two reasons: firstly, one of the vertices is clearly distinguished in their formulation. Secondly, Gauss' operators are typically horribly non-local. Nevertheless, in many cases it is possible to remove this problem by a canonical transformation. We will now discuss how to do this in general and then specialize to the case $\tau = \zeta$. 

Now suppose that $\tau$ is the boundary of a $2$-chain $\xi$. Let $F_{\xi}$ be the set of those $f \in F$ such that $(\xi ,f ) = 1$. For every $f \in F_{\xi}$ choose one vertex $v_f \in V$ incident to $f$. Define
\begin{equation}
\mathcal T v = \sum_{f \in F_{\xi}} \delta_{v,v_f} \cdot \partial f.
\end{equation}
Then one has $\partial \mathcal T=0$ and $\sum \limits_{v \in V} \mathcal T v = \tau$. Furthermore, $\mathcal Tv$ is at most the sum of some number of faces incident to the vertex $v$. Thus Gauss' operators are local and belong to the equivalence class specified by the cycle $\tau$.

\begin{figure}[H]
\includegraphics[scale=1] {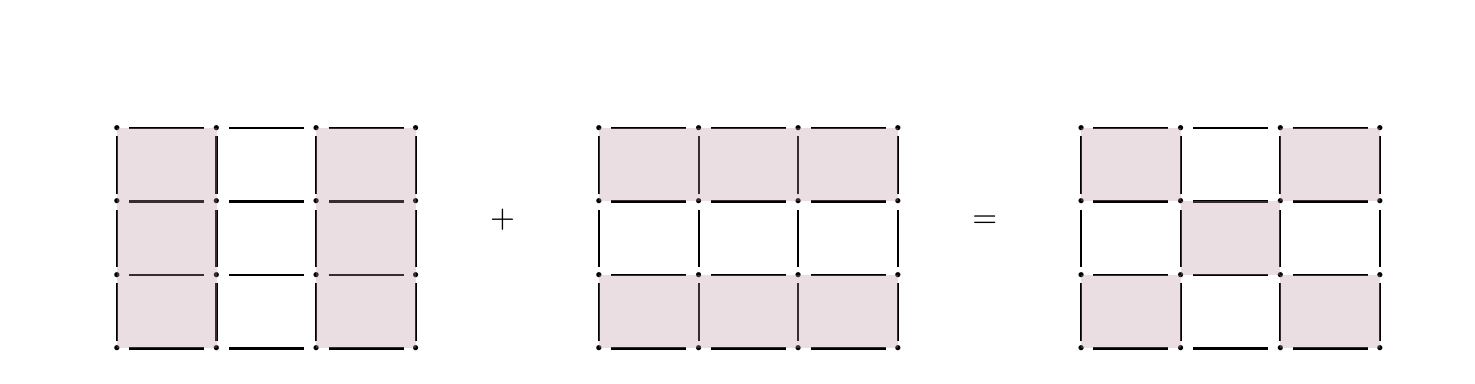}
\caption{Trivialization of $\zeta$ for a two-dimensional torus: $\zeta$ is the boundary of the sum of shaded faces, which are arranged in a pattern resembling a chessboard. Up to switching the roles of white and grey squares this is the only possibility in the two-dimensional case.}
\label{fig:pasiaki2D}
\end{figure}

\begin{figure}[H]
\centering
\includegraphics[scale=1] {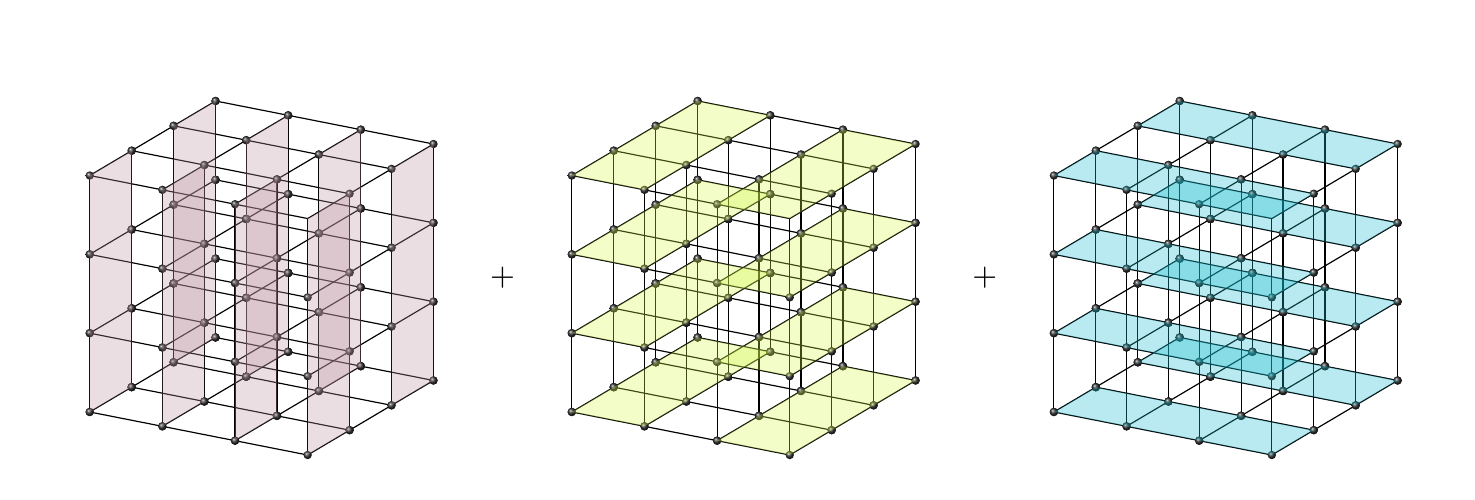}
\caption{Particular trivialization of $\zeta$ for a three-dimensional torus: $\zeta$ is the boundary of the sum of all colored faces. Taking all faces of one color only obtains the sum of all edges in one direction.}
\label{fig:pasiaki3D}
\end{figure}

The above discussion raises the question whether the outlined construction can be carried out for $\tau = \zeta$, leading to a local $\mathbb Z_2$ gauge theory equivalent to the $\Gamma$ model. Clearly this is always true for lattices representing simply-connected spaces, and more generally spaces $X$ such that the homology group $H_1(X, \mathbb Z_2)$ is trivial. Otherwise one has to ask whether $\zeta$ represents a nontrivial homology class. Interestingly, it is known \cite{toledo} that for a triangulation of a $d$-dimensional manifold $X$ which is obtained by barycentric subdivision of another triangulation, simplicial cycle $\zeta$ is Poincar\'{e} dual to the $(d-1)$-st Stiefel-Whitney class of $X$. However, the restriction to a very specific class of cell decompositions is important here. In general it is not possible to determine the homology class of $\zeta$ in terms of the topology of $X$ alone - it depends on the choice of decomposition. We will demonstrate this using the example of $d$-dimensional tori with arbitrary $d$. In this case all Stiefel-Whitney classes are trivial (since tori are parallelizable), but there exist decompositions for which $\zeta$ represents a nontrivial class, as well as such that $\zeta$ can be very explicitly trivialized. Indeed, for decompositions considered in subsection \ref{sec:toroidal}, cycle $\zeta$ is a~boundary if and only if at least two $L_i$ are even. If this condition is met, it is possible to construct trivializations of $\zeta$ invariant with respect to all translations by an even number of lattice sites. This is illustrated in figures \ref{fig:pasiaki2D} and \ref{fig:pasiaki3D} for dimensions two and three, respectively. 
Analogous construction works in any dimension. Using the notation of subsection \ref{sec:toroidal}, Gauss' operators in the two-dimensional take the form
\begin{equation}
    G(v) = (-1)^{(\eta,v)} \gamma(v) \cdot \begin{cases}
    U(\mathrm{NE}(v)) W(\delta v) & \text{for } $v$ \text{ even}, \\
    W(\delta v) & \text{for } $v$ \text{ odd},
    \end{cases}
\end{equation}
where $\eta$ is any $0$-chain with $\sum \limits_{v} (\eta,v)=\alpha$ and $\mathrm{NE}(v)$ is the plaquette to the north-east of $v$, i.e.\ the plaquette which has $v$ as its south-west corner.

\section{Duality with higher gauge theory} \label{sec:higher_gauge}

In this section we show how the $\Gamma$ model may be dualized to a $(d-1)$-form $\mathbb Z_2$ gauge theory, again with a modified Gauss' law. More precisely, we will dualize only operators $\Gamma_{\ast}(v)$ and $S(e)$, which generate the commutant of constraint operators arising in bosonization. Composition of this map with the correspondence between the $\Gamma$ model and fermions yields bosonization introduced in \cite{ck,ckr,chen}.

Hilbert space for the model we need is the tensor product of two-dimensional spaces associated to edges of the lattice. For each edge we introduce Pauli matrices $\left\{ \sigma_i(e) \right\}_{i=1}^3$. We~will think of $\sigma_3(e)$ as a higher dimensional parallel transport over a $(d-1)$-cell of the dual lattice\footnote{Strictly speaking it is not necessary to invoke the dual lattice to formulate this model. Indeed, all~formulas that follow will be written in terms of cells of the original lattice. However, their gauge theoretic interpretation is most directly seen on the dual lattice.}. As in ordinary gauge theory, parallel transports over individual cells will turn out not to be gauge-invariant. To construct an observable one has to take the product over all $(d-1)$-cells of some $(d-1)$-cycle. The simplest choice is the boundary of a $d$-cell, which corresponds to a vertex $v$ of the original lattice. This gives the operator
\begin{equation}
\mathsf{h}(v) = \prod_{e \ : \ v \in e} \sigma_3(e).
\end{equation}

Following \cite{ck,ckr,chen}, we would like to map operators $\Gamma_\ast(v)$ of the $\Gamma$ model to $\mathsf h(v)$. This is possible only upon restricting to the subspace defined by the condition $\prod \limits_{v \in V}\Gamma_{\ast}(v)=1$, because the product of all $\mathsf h(v)$ is equal to $1$ identically. In order to accommodate for existence of other states, we modify the mapping slightly by putting
\begin{equation}
\Gamma_{\ast}(v) \mapsto (-1)^{(\varepsilon, v)} \mathsf h(v)
\end{equation}
for some $0$-chain $\varepsilon$. We will think of $\varepsilon$ as a $d$-cochain on the dual lattice, or an external $d$-form field. What truly matters here is the quantity $\beta = \sum \limits_{v \in V} (\varepsilon, v)$, which characterizes the cohomology class of $\varepsilon$. Choices of $\varepsilon$ with the same $\beta$ give operator mapping related by conjugation with the product of some number of $\sigma_1(e)$ operators.

Next we construct an operator corresponding to $S(e)$. We consider the Ansatz
\begin{equation}
S(e) \mapsto \mathsf e(e) = \sigma_1(e) \cdot \prod_{e'} \sigma_3(e')^{\nu(e,e')}
\label{eq:S_to_l}
\end{equation}
for some function $\nu : E \times E \to \mathbb Z_2$. Since $S(e)^2=-1$, we need to have $\mathsf e(e)^2=-1$. This yields the condition $\nu(e,e)=1$, which will be assumed from now on.

With the above definitions, braiding relations between $\mathsf h(v)$ and $\mathsf e(e)$ operators are correct, but we still need to impose braiding relations between distinct $\mathsf e(e)$. The sought-after condition is $\mathsf e(e) \mathsf e(e') = (-1)^{(\partial e, \partial e')} \mathsf e(e') \mathsf e(e) $, which translates to
\begin{equation}
 \nu(e,e') + \nu(e',e) = (\partial e, \partial e'). 
 \label{eq:nu_def}
\end{equation}
If edges $e, e'$ do not share a common vertex, the above relation asserts that $\nu(e,e')=\nu(e',e)$. It seems to be most natural to put
\begin{equation}
\nu(e,e') = \nu(e',e) = 0 \qquad \text{if} \quad (\partial e, \partial e')=0, \ \ e \neq e'.
\label{eq:nu_loc}
\end{equation}
With this requirement, operators $\mathsf e(e)$ consist only of Pauli matrices on edges $e'$ in the nearest vicinity of $e$, assuring that the mapping is local. On the other hand, if $e$ and $e'$ share one common vertex, values of $\nu(e,e')$ and $\nu(e',e)$ have to be opposite. In other words, we have to choose either $\nu(e,e')=1$ and $\nu(e',e)=0$, or vice versa. 

In the above we have argued that functions $\nu$ satisfying \eqref{eq:nu_def} exist and may be subjected to the additional locality condition \eqref{eq:nu_loc}. Next, we will demonstrate that it is essentially unique, in the sense that distinct choices yield operator maps related by a local unitary rotation. Indeed, given two such functions $\nu_1, \nu_2$ we put $\omega(e,e') = \nu_1(e,e')+\nu_2(e,e')$. Then $\omega(e,e)=0$ and $\omega(e,e')=\omega(e',e)$. Transformation
\begin{equation}
\sigma_1(e) \mapsto \sigma_1(e) \cdot \prod_{e'} \sigma_3(e')^{\omega(e,e')} , \qquad \sigma_3(e) \mapsto \sigma_3(e)
\end{equation}
defines an algebra automorphism, see appendix \ref{canonical_heisenberg}. It is local if both $\nu_1$ and $\nu_2$ satisfy \eqref{eq:nu_loc}. By construction, it is such that
\begin{equation}
\mathsf h(v) \mapsto \mathsf h(v), \qquad \sigma_1(e) \cdot \prod_{e'} \sigma_3(e')^{\nu_1(e,e')} \mapsto \sigma_1(e) \cdot \prod_{e'} \sigma_3(e')^{\nu_2(e,e')}.
\end{equation}
In a similar way, any signs that could be included in the definition of $\mathsf e(e)$ could also be reabsorbed by a unitary transformation, so we do not consider including them. 
Besides the braiding relations, there exist certain global constraints that have to be taken into account. We have already mentioned that the operator $\Gamma_{\ast} = \prod \limits_{v \in V} \Gamma_{\ast}(v)$ is sent to the c-number $(-1)^{\beta}$, so this mapping may be valid only upon restricting to the corresponding subspace of the $\Gamma$ model. On the other hand, by \eqref{eq:fermi_parity_A_rel}, we have that on this subspace $S(\ell)=(-1)^{\alpha + \beta}$ for an Eulerian circuit $\ell=(e_1,...,e_n)$. This gives a constraint
\begin{equation}
\mathsf e(e_1) \ldots \mathsf e(e_n) = (-1)^{\alpha + \beta},
\end{equation}
which is a nontrivial restriction, since the left hand side is not a c-number. It is not difficult to check that this equation is satisfied on a subspace of dimension $2^{|E|-1}$, which is also equal to the dimension of the Hilbert space of the $\Gamma$ model with fixed value of $\Gamma_{\ast}$.

Using methods and results of previous sections, it is easy to check that there are no further independent relations satisfied by operators $\Gamma_{\ast}(v)$ and $S(e)$. Therefore, the proposed map well-defines a homomorphism of operator algebras.

Construction of the duality is now essentially completed. We will now give the correspondence between various notions formulated in the two pictures of the model.

Firstly, for every circuit $\ell = (e_1,...,e_n)$ the operator $S(\ell)$ is mapped to a certain operator $\mathsf e(\ell)$. Its explicit form is easy to evaluate:
\begin{equation}
\mathsf e(\ell) = \left(-1\right)^{\sum \limits_{i < j} \nu(e_i, e_j)} \cdot \prod_{i=1}^n \sigma_1(e_i) \cdot \prod_{e'} \sigma_3(e')^{\sum \limits_{i=1}^n \nu(e_i, e')}.
\label{eq:e_op_explicit}
\end{equation}
This operator commutes with all $\mathsf h(v)$ and $\mathsf e(e)$ and satisfies $\mathsf e(\ell)^2=1$. In the case that $\ell$ is the loop around the boundary of a face $f$, we interpret $\mathsf e(\ell)$ as a Gauss' operator of the gauge theory and denote it by $\mathsf g(f)$. Indeed, it satisfies the expected relation
\begin{equation}
\mathsf g(f) \sigma_3(e) = (-1)^{(\partial f, e)} \sigma_3(e) \mathsf g(f).
\end{equation}
States violating the constraint $S(\ell)=1$ were interpreted earlier in terms of fermions propagating in an external $\mathbb Z_2$ gauge field $A$. On the higher gauge theory side this background field is thought of as a $(d-2)$-form electric charge distribution, localized on the cycle Poincar\'{e} dual to $\delta A$. Indeed, for these states we have $\mathsf g(f) = (-1)^{(\delta A,f)}$.

Operators $\mathsf e(\ell)$ with $\ell$ non-contractible furnish a $(d-1)$-form $\mathbb Z_2$ symmetry of the higher gauge theory. They act trivially on all $\mathsf h(v)$, but not so on holonomies along homologically nontrivial $(d-1)$-cycles of the dual lattice. Flat background gauge fields for fermions correspond to eigenspaces of this symmetry. The symmetry is subject to a 't Hooft anomaly, whose form was identified in \cite{dk}. One may understand the presence of the anomaly in an intuitive way as follows. We see from \eqref{eq:e_op_explicit} that the symmetry is not on-site (the correct notion of a site being an edge, i.e.\ a dual $(d-1)$-cell). Thus there is no canonical gauging procedure, but one can attempt to take a different route. Symmetry operators $\mathsf e(\ell)$ furnish a representation of the group of $1$-cycles (dual $(d-1)$-cocycles). We would like to extend it to an action of the group all $1$-chains (dual $(d-1)$-cochains). The simplest choice would be to take the operator corresponding to the transformation at the edge $e$ to be $\mathsf e(e)$, but this does not work, since $\mathsf e(e)$ placed at different edges commute only up to signs. It is expected that there is no way around this difficulty.

We remark that in \cite{ck,ckr,chen} a more specific construction of duality between fermions and higher gauge theory, essentially corresponding to a particular choice of the function $\nu$, was presented. It is formulated in terms of higher cup products \cite{steenrod} and depends on a choice of a branching structure\footnote{Branching structure is a choice of orientations of edges such that there is no loop in any triangle.} on the dual lattice, assumed to be a triangulation. An interesting feature of this approach is that certain sign factors appearing in constraint operators may be expressed in terms of a cocycle $w_2$ representing the second Stiefel-Whitney class. Given a spin structure, understood as a cochain $E$ such that\footnote{In the notation of papers we are refering to, this equation takes the form $\delta E = w_2$, but in the two perspectives the roles of the lattice and its dual are reversed, so operators $\delta$ and $\partial$ are exchanged.} $\delta E = w_2$, one may absorb these signs by redefining fermionic kinetic operators. The fact that signs may be shuffled between the definition of the bosonization map and the form of Gauss' operator may be traced to the duality between background gauge fields and background electric charge distributions. We remark that the role of spin structures in bosonization has been discussed also in \cite{rad_spin}.

Despite the elegance of the construction outlined above, we would like to emphasize that existence of a spin structure is not necessary to construct bosonization maps. In fact all models considered in this paper make sense on a large class of graphs, which are not necessarily discretizations of manifolds and hence do not have well-defined Stiefel-Whitney classes or spin structures. Of course, it may very well be true that spin structures do play an indispensable role if one imposes some \textit{naturality} or \textit{functoriality} conditions on duality maps, but it is not completely clear to us what would be the correct formulation of this. On the other hand, spin structures clearly become important in concrete dynamical models. For example, continuum limits of many lattice models with fermions should depend on a~spin structure. Another interesting example of this is the discussion of the Gu-Wen model \cite{guwen} in \cite{dk}. 

\section{Summary and outlook\label{sec:summary}}

We generalized the bosonization prescription based on the $\Gamma$ model, presented a new proof of its correctness and reinterpreted it in terms of lattice $\mathbb Z_2$ gauge theory. We found that its alternative gauge-theoretic description involves modified Gauss' operators, much as in Chern-Simons-like theories. Furthermore, we discussed the duality with higher gauge theories recently proposed in the context of bosonization. These results are valid independently of the spatial dimension.

In order to actually perform bosonization (rather than to couple fermions to gauge fields) it is necessary to introduce constraints in the $\Gamma$ model. They can be interpreted as a~flatness condition for the gauge field. We have presented a solution of these constraints in the case of two-dimensional tori. Unfortunately, our method does not seem to generalize to higher dimensions in a straightforward way. Thus dealing with constraints in an efficient way for general geometries remains a challenge for future research.

Another interesting problem not solved for now is to obtain a useful state-sum formulation of the $\Gamma$ model. Furthermore, it remains to be seen whether it is possible to apply constructions of this type to shed new light on some problems in lattice gauge theory, such as those related to fermion doubling or anomalous symmetries.

\appendix 

\section{Canonical transformations for Ising degrees of freedom \label{canonical_heisenberg}}

In this appendix we summarize properties of the Heisenberg groups for $\mathbb Z_2$-valued degrees of freedom and their automorphisms, called canonical transformations. There are essentially no new results here, but we do not know a reference in which the whole material presented here is discussed concisely. We refer to \cite{weil_01} and \cite{blasco} for further discussion.

Let $M$ be a finite-dimensional $\mathbb Z_2$-vector space. A bilinear form $\Omega : M \times M \to \mathbb Z_2$ is said to be alternating if $\Omega(x,x)=0$ for every $x \in M$. Every alternating form is symmetric, but the converse is not true\footnote{It is true over any field that alternating forms are skew-symmetric, but in the case of fields of characteristic two skew-symmetry and symmetry is the same. Furthermore, it is true in general that skew-symmetry of a form $\Omega$ implies that $2 \Omega(x,x)=0$ for every $x \in M$. This implies that $\Omega$ is alternating if $2$ is invertible, but it is a~vacuous statement in the case of characteristic two.}. Alternating form which is non-degenerate, i.e.\ such that for every $x \in M$ there exists $y \in M$ such that $\Omega(x,y)=1$, is called a symplectic form. If $\Omega$ is a~symplectic form, there exists a basis $\{ e_i, f_i \}_{i=1}^n$ in which $\Omega$ takes the canonical form
\begin{equation}
\Omega(e_i,e_j)= \Omega(f_i,f_j)=0, \qquad \Omega(e_i,f_j) = \delta_{i,j}.
\label{eq:symplectic_canonical}
\end{equation}
In particular, the dimension of $M$ is necessarily even. It is the only invariant of $(M, \Omega)$.

Function $Q : M \to \mathbb Z_2$ is called a quadratic form if the map $\Omega : M \times M \to \mathbb Z_2$ given by $\Omega(x,y) = Q(x+y)-Q(x)-Q(y)$ is a bilinear form. Bilinear forms $\Omega$ arising this way are automatically alternating. If $\Omega$ is also non-degenerate, we say that $Q$ is non-singular. We~assume this condition from now on. Thus $\dim(M) = 2n$. Subspace $N \subseteq M$ is said to be isotropic if $Q(x)=0$ for every $x \in N$. One can show that maximal isotropic subspaces of $M$ are of dimension $n$ or $n-1$. These two possibilities correspond to values $0$ and $1$, respectively, of the so called Arf invariant $\mathrm{Arf}(Q)$ of $Q$ \cite{arf}. Dimension of $M$ and the Arf invariant are the only invariants of $(M,Q)$. In the case $\mathrm{Arf}(Q)=0$ it is possible to choose a basis in which $\Omega$ takes the canonical form \eqref{eq:symplectic_canonical} and additionally $Q(e_i)=Q(f_i)=0$.

Let $(M,Q)$ be as in the previous paragraph and let $\mathfrak B = \{ x_i \}_{i=1}^{2n}$ be a basis of $M$. The~Heisenberg group $H_{Q,\mathfrak B}$ is the group with generators $\{ z \} \cup \{ T_i \}_{i=1}^{2n}$, subject to relations
\begin{equation}
z^2 = 1, \qquad T_i^2 = z^{Q(x_i)}, \qquad zT_i = T_i z, \qquad T_i T_j = z^{\Omega(x_i,x_j)} T_j T_i.
\end{equation}
Its center $Z(H_{Q, \mathfrak B})$ is generated by the element $z$. Quotient $H_{Q, \mathfrak B}/Z(H_{Q, \mathfrak B})$ is a $\mathbb Z_2$-vector space. It may be identified with $M$, with the coset of $T_i$ corresponding to the element~$x_i$. We~let $\pi$ be the canonical projection $H_{Q, \mathfrak B} \to M$. It is easy to check that $gg'=z^{\Omega(\pi(g),\pi(g'))} g'g$ and $g^2 = z^{Q(\pi(g))}$ for every $g,g' \in H_{q, \mathfrak B}$.

Suppose $M'$ is another $\mathbb Z_2$-vector space and let $\varphi : (M,Q) \to (M',Q')$ be an isometry, i.e.~a~linear map such that $Q'(\varphi(x))=Q(x)$ for every $x \in M$. Choose a basis $\mathfrak B'$ of $M'$ and consider the group $H_{Q', \mathfrak B'}$. For each $x_i$ we can find a (non-unique) $T_i' \in H_{Q', \mathfrak B'}$ such that $\pi(T_i')=\varphi(x_i')$. Elements $T_i'$ satisfy all relations obeyed by $T_i$, so there is a unique group homomorphism $ \Phi : H_{Q, \mathfrak B} \to H_{Q', \mathfrak B'}$ such that
\begin{equation}
\Phi(z) = z, \qquad \Phi(T_i) = T_i'.
\end{equation}
Clearly $\Phi$ is a lift of $\varphi$, in the sense that $\pi \circ \Phi = \varphi \circ \pi$. We emphasize that homomorphisms $ \Phi$ lifting $\varphi$ are not unique, because in the above constructions we have to choose elements $T_i'$, with distinct choices corresponding to distinct lifts. This means that Heisenberg groups corresponding to $(M,Q)$ constructed using different bases of $M$ are isomorphic, but not canonically isomorphic\footnote{There is a canonically defined class of isomorphisms modulo compositions with inner automorphisms.}. Having said that, we will abuse the notation slightly by abbreviating $H_{Q,\mathfrak B}$ to $H_{Q}$.

In this paper we will use only quadratic forms $Q$ with $\mathrm{Arf}(Q)=0$. In this case we can choose a basis of $M$ in which $Q$ takes the canonical form. We let $U_i, W_i \in H_Q$ be some lifts of $e_i$ and $f_i$, respectively.

Automorphisms of $H_Q$ will be called canonical transformations. Every canonical transformation $\Phi$ acts trivially on $Z(H_Q)$, hence induces an isometry $\varphi$ of $(M,Q)$. The map $\widetilde \pi : \Phi \mapsto \varphi$ is a homomorphism from $\mathrm{Aut}(H_Q)$ to $\mathrm{O}(M,Q)$, the orthogonal group of $Q$. It~is~clear from the preceding discussion that $\widetilde \pi$ is surjective. Next, let $\Phi$ be in the kernel of~$\widetilde \pi$. Then we have $ \pi \circ \Phi(U_i)= e_i$ and $ \pi \circ \Phi(W_i)=f_i$, so
\begin{equation}
\Phi(U_i)=(-1)^{a_i} U_i, \qquad \Phi(W_i) = (-1)^{b_i} W_i 
\end{equation}
for some $a_i, b_i \in \mathbb Z_2$. Conversely, for every collection $\{ a_i, b_i \}_{i=1}^n$ the above formula defines a~canonical transformation $\Phi \in \mathrm{ker}(\widetilde \pi)$. Automorphisms of this form are precisely the inner automorphisms: $\Phi(g') = gg'g^{-1}$ with $g = \prod\limits_{i=1}^n U_i^{a_i} W_i^{b_i} \in H_q$. Therefore $\mathrm{ker}(\widetilde \pi)$ may be identified with $M$, since $Z(H_Q)$ is precisely the group of those elements of $H_Q$ which act trivially on $H_Q$. We have shown that $\mathrm{Aut}(H_Q)$ is an extension of $\mathrm{O}(M, Q)$ by $M$. Interestingly, it~is known that this extension is non-split for $n \geq 3$ \cite{griess}. This is in contrast with the more standard situation for Heisenberg groups in characteristic different than two.

The last issue we need to touch upon is representation theory. First we define the standard representation $\rho$ of $H_Q$ on $\left( \mathbb C^2 \right)^{\otimes n}$ by
\begin{equation}
\rho(z) = -1, \qquad \rho(U_i) = 1_{\mathbb C^2}^{\otimes (i-1)} \otimes \sigma_3 \otimes 1_{\mathbb C^2}^{\otimes (n-i)}, \qquad \rho(W_i) = 1_{\mathbb C^2}^{\otimes (i-1)} \otimes \sigma_1 \otimes 1_{\mathbb C^2}^{\otimes (n-i)},
\end{equation}
where $\{ \sigma_i \}_{i=1}^3$ are the Pauli matrices. It is easy to see that this representation is irreducible. We claim that up to isomorphism this is the only irreducible representation of $H_Q$ on which $Z(H_Q)$ acts nontrivially. Indeed, representations on which $Z(H_Q)$ acts trivially are in one-to-one correspondence with representations of $M$. Now recall \cite{kirillov} that the number of non-isomorphic irreducible representations of a finite group is equal to the number of its conjugacy classes. It is easy to check that the number of conjugacy classes in $H_Q$ exceeds the number of conjugacy classes in $M$ by one, which completes the argument. The statement just proven is an analogue of the Stone -- von Neumann theorem for Ising degrees of freedom. It has an additional corollary that every non-trivial normal subgroup of $H_Q$ contains $z$.

Now let $\Phi$ be a canonical transformation. Then $\rho \circ \Phi$ is also an irreducible representation on which the center acts nontrivially, so by the above theorem there exists a unitary endomorphism $p(\Phi)$ of the standard module, unique up to phase, such that $\rho (\Phi(g)) = p(\Phi) \rho(g) p(\Phi)^{-1}$ for every $g \in H_Q$. Assignment $\Phi \mapsto p(\Phi)$ is a projective representation of the group of canonical transformations. Even its restriction to $M \subseteq \mathrm{Aut}(H_Q)$ is not equivalent to a linear representation. It can be lifted to a genuine representation of a certain finite central extension of $\mathrm{Aut}(H_Q)$. The structure of this extension is not known to us, but fortunately it will not be needed. The important point is the existence of $p$.

\section{Graphs with vertices of odd degree \label{sec:odd_degree}}

In this appendix we shall briefly describe a generalization of the $\Gamma$ model to the case in which some vertices have odd degree. It will be shown that this has the effect of introducing additional degrees of freedom on each vertex of odd degree. We construct operators which create and annihilate these excitations.

We decompose the set of vertices $V$ into two disjoint sets $V_{\alpha}$ of vertices of degree $\alpha\pmod{2}$. Hilbert spaces associated to vertices of even degree are constructed as earlier. For~vertices $v$ of odd degree the Clifford algebra generated by $\{ \Gamma_\ast(v) \} \cup \{ \Gamma(v,e) \}_{e \in \mathrm{St}(v)}$ has one (rather than two) non-isomorphic irreducible representation. We take this representation as the Hilbert space associated to $v$. In contrast to the even case, $\Gamma_\ast(v)$ cannot be expressed in terms of other generators. With this definition, the dimension of the full Hilbert space $\mathcal{H}$ is 
\begin{equation}
    \dim(\mathcal{H})=\prod\limits_{v\in V_0} 2^{\frac{\deg(v)}{2}}\prod\limits_{v\in V_1}2^{\frac{\deg(v)+1}{2}}=2^{|E|+\frac{1}{2}|V_1|}.
\end{equation}
Since this is an integer, it follows that $|V_1|$ is even. This can also be seen by reducing the equation $\sum \limits_{v \in V} \deg(v) = 2 |E|$ modulo two. 

As in the case of graphs with even vertices only, we can decompose $\mathcal H = \bigoplus \limits_{[A] \in Z_1^\ast} \mathcal H_{[A]}$. Calculation analogous to the one in equation (\ref{eq:O_operators}) shows that each $\mathcal H_{[A]}$ has the same dimension. Since the number of distinct $[A]$ is $2^{|E|-|V|+1}$, one has 
\begin{equation}
    \dim(\mathcal{H}_{[A]})=\frac{2^{|E|+\frac{1}{2}|V_1|}}{2^{|E|-|V|+1}}=2^{|V|-1}\cdot 2^{\frac{1}{2}|V_1|}.
\end{equation}
Thus $\mathcal{H}_{[A]}$ is as large as $2^{\frac{1}{2}|V_1|}$ "halves" of the Fock space. To account for this multiplicity we introduce new operators. Let $\ell=(e_1,...,e_n)$ be a path with initial point $ v= s(e_1)$ and final point $v' =t(e_n)$ of odd degrees. We define
\begin{equation}
    \Psi(\ell)= i^{\frac{\deg(v)+\deg(v')}{2}+1} \left(\Gamma_\ast(v) \cdot \prod\limits_{e\in \mathrm{St}(v)}\Gamma(v,e)\right)  S(\ell)\left(  \Gamma_\ast(v') \cdot \prod\limits_{e\in \mathrm{St}(v')}\Gamma(v',e)  \right),
\end{equation}
where we choose some orderings of $\mathrm{St}(v)$ and $\mathrm{St}(v')$, modulo even permutations. 

We list in points the main properties of $\Psi(\ell)$:
\begin{itemize}
    \item $\Psi(\ell)$ commutes with $S(\ell')$ and $\Gamma_*(v'')$ for any path $\ell'$ and any vertex $v''$. 
    \item If $\ell'$ is a path with initial vertex $v'$ and final vertex $v''$, then
    \begin{equation}
    \Psi(\ell)\Psi(\ell') = 
    \begin{cases}
    \Psi(\ell \ell') & \text{if } v \neq v'', \\
    S(\ell \ell') & \text{if } v = v'',
    \end{cases}
    \end{equation}
    where $\ell \ell'$ is the concatenation of $\ell$ and $\ell'$.
    \item We have braiding relations
    \begin{equation}
        \Psi(\ell) \Psi(\ell') = (-1)^{(\partial [\ell],\partial [ \ell' ])} \Psi(\ell') \Psi(\ell).
    \end{equation}
    \item $\Psi(\ell)^2=-1$.
\end{itemize}
One can further decompose each $\mathcal H_{[A]}$ into subspaces corresponding to even and odd numbers of fermions, $\mathcal H_{[A],0}$ and $\mathcal H_{[A],1}$. One can show that each $\mathcal H_{[A],\alpha}$ is an irreducible representation (of dimension $2^{|V| + \frac{1}{2}|V_1|-2}$) of the algebra $\mathcal A_0 \otimes_{\mathbb C} \mathbb C \mathcal G$, where $\mathbb C \mathcal G$ is the group algebra of the group $\mathcal G$ generated by all $\Psi$ operators.

There are some similarities between the presented structure and the so-called delocalized fermions \cite{kitaev,topCM}, considered e.g.\ in the field of topological quantum computation. These excitations consist of multiple fermionic degrees of freedom, located at different lattice sites and connected by strings.

\acknowledgments

Initial stage of this project has been realized with J.~Wosiek and A.~Wyrzykowski. We~thank J.~Wosiek for an introduction to the subject of bosonization, discussions and encouragement. We~are grateful to Y-A.~Chen, A.~Francuz, L.~Hadasz, Z.~Komargodski, M.~Rocek, K.~Roumpedakis and S.~Seifnashri for discussions. Analysis carried out in section \ref{sec:higher_gauge} has been suggested to us by an anonymous referee. BR~was supported by the NCN grant UMO-2016/21/B/ST2/01492 and the MNS donation for PhD students and young scientists N17/MNS/000040.

\end{document}